\documentclass[Afour,sagev,times]{sagej}

\usepackage{moreverb,url}

\usepackage[colorlinks,bookmarksopen,bookmarksnumbered,citecolor=red,urlcolor=red]{hyperref}

\usepackage[table]{xcolor} 
\usepackage{multirow}
\usepackage{enumitem}
\usepackage{array}
\newcolumntype{L}[1]{>{\raggedright\let\newline\\\arraybackslash\hspace{0pt}}m{#1}}
\newcolumntype{C}[1]{>{\centering\let\newline\\\arraybackslash\hspace{0pt}}m{#1}}
\newcolumntype{R}[1]{>{\raggedleft\let\newline\\\arraybackslash\hspace{0pt}}m{#1}}

\newcommand\BibTeX{{\rmfamily B\kern-.05em \textsc{i\kern-.025em b}\kern-.08em
T\kern-.1667em\lower.7ex\hbox{E}\kern-.125emX}}

\def\volumeyear{2021}

\begin{document}

\runninghead{Keith, Mitra, and North}

\title{Design guidelines for narrative maps in sensemaking tasks}

\author{Brian Felipe Keith Norambuena \affilnum{1,3}, Tanushree Mitra \affilnum{2}, and Chris North \affilnum{1}}

\affiliation{\affilnum{1}Virginia Tech, USA\\
\affilnum{2}University of Washington, USA\\
\affilnum{3}Universidad Católica del Norte, Chile}

\corrauth{Brian Keith Norambuena, 
InfoVis Lab, Sanghani Center for Artificial Intelligence \& Data Analytics
Virginia Tech,
Blacksburg, Virginia,
24061, USA.}

\email{briankeithn@vt.edu}

\begin{abstract}
Narrative sensemaking is a fundamental process to understand sequential information. Narrative maps are a visual representation framework that can aid analysts in their narrative sensemaking process. Narrative maps allow analysts to understand the big picture of a narrative, uncover new relationships between events, and model the connection between storylines. We seek to understand how analysts create and use narrative maps in order to obtain design guidelines for an interactive visualization tool for narrative maps that can aid analysts in narrative sensemaking. We perform two experiments with a data set of news articles. The insights extracted from our studies can be used to design narrative maps, extraction algorithms, and visual analytics tools to support the narrative sensemaking process. The contributions of this paper are three-fold: (1) an analysis of how analysts construct narrative maps; (2) a user evaluation of specific narrative map features; and (3) design guidelines for narrative maps. Our findings suggest ways for designing narrative maps and extraction algorithms, as well as providing insights towards useful interactions. We discuss these insights and design guidelines and reflect on the potential challenges involved. As key highlights, we find that narrative maps should avoid redundant connections that can be inferred by using the transitive property of event connections, reducing the overall complexity of the map. Moreover, narrative maps should use multiple types of cognitive connections between events such as topical and causal connections, as this emulates the strategies that analysts use in the narrative sensemaking process.
\end{abstract}


\maketitle

\section{Introduction}
Narratives are systems of stories \cite{halverson2011master}---sequences of events tied together in a coherent fashion. Events are the fundamental units of narrative action, they are either an act involving characters and entities or a happening where no entities are causally involved \cite{abbott2008cambridge}. Narratives are fundamental to our understanding of the world and provide a natural way to capture relationships between sequences of events, as well as the goals, motivations, and plans of actors \cite{finlayson2013military}. Narratives are used in the process of ``connecting the dots" between apparently unrelated pieces of information \cite{hossain2011helping, hossain2012connecting} and modeling causal relationships \cite{choudhry2020once}.

Storytelling in general is an accepted metaphor used in visual analytics and analytical reasoning \cite{segel2010narrative,tong2018storytelling,riche2018data}. However, unlike general visual storytelling, our work focuses specifically on visualizing textual narratives, such as those created by news. In this context, narratives provide a way to understand the information landscape, a key part of several narrative sensemaking tasks \cite{keith2020maps}. Example narrative sensemaking tasks range from a journalistic analysis of news narratives \cite{bradel2015big}, where the goal might be to understand the big picture, to intelligence analysis  \cite{endert2014human}, where the goal is to uncover hidden or implicit relations between events. 

To aid analysts with sensemaking tasks, scholars have created visual analytics software, which allow analysts to process and understand greater quantities of data and information \cite{cook2005illuminating}. These tools focus on different parts of the sensemaking loop \cite{pirolli2005sensemaking}. For example, while some tools focus on the foraging loop \cite{kang2009evaluating}, others focus on the synthesis loop \cite{wright2006sandbox} to generate hypotheses. However, there is still a lack of support towards building tools that use narrative representations to aid in narrative sensemaking tasks, such as connecting events, extracting storylines, and constructing narratives \cite{keith2020maps}.

In this work, we focus on a specific type of graph-based visual narrative representation---narrative maps \cite{keith2020maps}. Narrative maps are a specific type of a narrative graph representation that uses events as its representational basic unit. provide a generic foundation to encode different types of narratives extracted from data, requiring only the existence of a total ordering (e.g., in the form of timestamps) and text representation of the event (e.g., a news headline). Narrative maps are a useful visualization framework to understand the information landscape. As a sensemaking tool, narrative maps have applications in intelligence analysis, misinformation modeling, and computational journalism \cite{keith2020maps}. In particular, they offer a way to keep track of the big picture of a narrative in the context of the ever-increasing problem of information overload \cite{ho2001towards,shahaf2010connecting}. Moreover, they allow for uncovering connections between events in the narrative, which helps analysts connect the dots and understand events as well as their context. Furthermore, narrative maps could be used to explore how narratives and counter-narratives emerge over time, thus providing a way to model how misinformation spreads \cite{keith2020maps}. 

However, from a visualization standpoint, the optimal design of narrative maps for the sensemaking process remains unexplored. We attempt to remedy this gap by defining a series of \textit{design guidelines} for narrative maps. In particular, we explore how analysts create, structure, and use narrative maps to determine the characteristics of \textit{good} narrative maps. Through our exploration, we develop design guidelines that provide the basis for the creation of an interactive visualization toolkit for narrative maps; this toolkit can aid analysts in their narrative sensemaking process. Thus, the contributions of our paper are the following: (1) an analysis of how analysts construct narrative maps, including the types of cognitive connections and structures; (2) a user evaluation of specific narrative map features, namely size and transitivity; (3) a series of design guidelines for narrative maps and extraction algorithms.

Finally, the overarching goal in this work is to improve the design of narrative maps and their associated extraction algorithms \cite{keith2020maps}. Narrative maps made heavy use of narrative theory in their inception, but its original design did not include analyst feedback in the context of the narrative sensemaking process. Thus, the main findings and design guidelines proposed in this article provide empirical scaffolding in the context of sensemaking that can be used to improve the design of narrative maps and their associated extraction algorithms.

The rest of this article is organized as follows. First, we present a motivating example about narrative maps, which leads to the two research questions explored in this work. Afterward, we discuss related work on narrative visualization, extraction, and representation, as well as previous work studying cognitive strategies in the sensemaking process. Then, we present our empirical study on narrative map construction for sensemaking (RQ1), showcasing the different strategies used by analysts. Then, we discuss the specific effects of using connections that can be induced by transitivity and the size of the map through a user evaluation (RQ2). Using both results, we present a series of design guidelines. Next, we present an in-depth discussion of our results and their implications to the sensemaking process. Finally, we present the conclusions of our work.

\section{Narrative Maps}
\subsection{Motivating Example}
To show how narrative maps work, consider the narrative surrounding the Coronavirus outbreak at the start of 2020 using real data extracted from news articles. Bob, an analyst working in investigative journalism, wants to explore how the start of the outbreak led to the US travel restrictions. Moreover, he is interested in exploring other outcomes of the outbreak during this time. These two tasks are examples of narrative sensemaking. In particular, finding out how two events are connected is a \textit{directed task}, because the analysis is focused on understanding the connection between the two events. In contrast, exploring all the outcomes of the outbreak is an \textit{open-ended task}, as it does not focus on any particular outcome, leaving the analyst with more room to explore the branching system of stories. Thus, Bob decides to use a narrative map with a data set of articles on the Coronavirus outbreak from the top five news sources at the time. We will show how these two questions can be answered using a narrative map. 

In general, narrative maps can be used to answer the directed and open-ended tasks \cite{keith2020maps}. That is, their main purpose is to aid analysts in connecting the dots between events, such as those represented by news articles or intelligence reports, and understanding the different storylines that emerge from these events. Thus, narrative maps provide a generic sensemaking framework for analysts. In particular, intelligence analysts could also use it as a graphical representation of their mental model, similar to other narrative-based models \cite{bex2006anchored,sappelsa2013generic}.

\begin{figure}[!htb]
    \centering
    \includegraphics[width=\columnwidth]{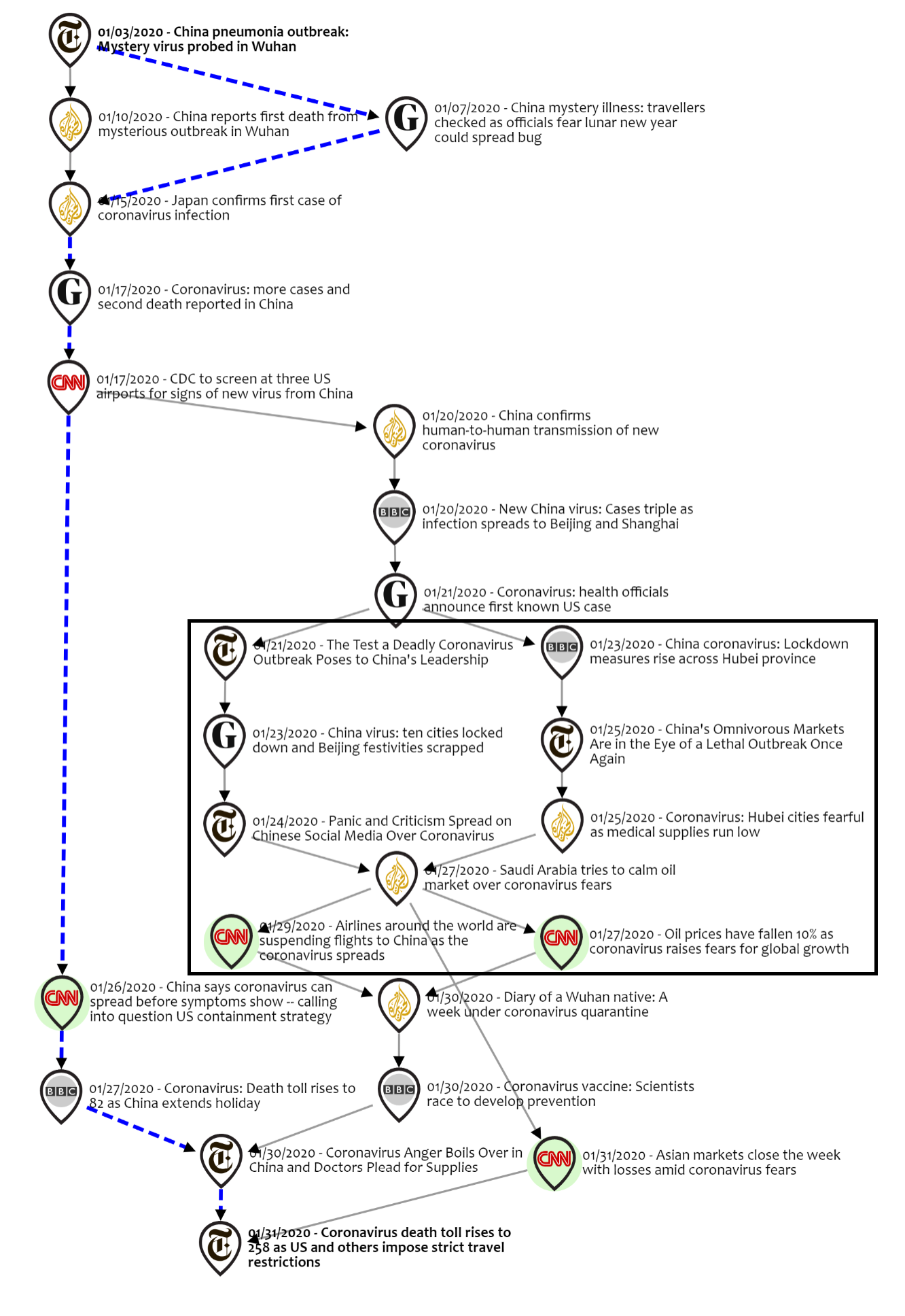}
    \caption{Example of a Narrative Map showing the COVID-19 narrative in January 2020 from news articles. The highlighted panel shows some important outcomes of the outbreak (lockdown in China, social effects, and economic effects).}
    \label{fig:example}
\end{figure}

In this context, Bob selects two points of interest based on his tasks: a starting and ending point for the narrative. In particular, he starts the narrative with the mysterious pneumonia outbreaks in Wuhan at the start of the month and ends it with the US imposing travel restrictions. The extraction algorithm selects a coherent subset of these articles to build a visual representation of the underlying narrative. We show the output visualization in Figure \ref{fig:example}. 


After extracting the narrative, we find the main storyline---the most coherent path in the graph---which we represent with dashed blue edges. Next, we find the important events---a set of representative events from each storyline---which we highlight with green nodes. These events give us an overview of the side storylines of the narrative and focus on issues not covered by the main storyline. 

To complete the directed task, Bob looks at the main storyline, which begins with the mysterious outbreak. 
Based on the main story of the narrative, Bob is able to identify the core causes of imposing travel restrictions: rising cases and deaths, medical supply issues, and asymptomatic spread.

To complete the open-ended task, Bob looks at the side storylines. In particular, he focuses on the zoomed-in section of the map. This area shows some key side storylines. 
Bob is able to identify three important outcomes from the narrative map: lockdowns, social impacts, and economic impacts.

\subsection{Research Questions}
The motivating example shows how an analyst could apply a narrative map to extract important information from the data. Studying the narrative map allowed the analyst to answer the questions defined by the directed and open-ended tasks. Having shown this narrative map example, we now present our research questions. As mentioned previously, our goal is to determine the characteristics of a \textit{good} narrative map. We do this by understanding how analysts construct narrative maps, as this gives us an insight into the structures and types of connections they would use, and we also explore how specific characteristics affect the utility of narrative maps from a consumer perspective. Thus, we sought to answer the following research questions:

\begin{itemize}
    \item \textbf{RQ1: How do analysts manually construct narrative maps?}
    We focus on the strategies, cognitive connections, and structures used during the construction process of the narrative map. 
	\item \textbf{RQ2: How do map size and transitivity affect the utility or effectiveness of the map?}
	We explore the effects of different combinations of map size---as defined by the length of the main story---and transitivity---whether we should allow redundant connections that can be induced by transitivity or not. 
\end{itemize}

\begin{figure}[!htb]
    \centering
    \includegraphics[width=0.5\textwidth]{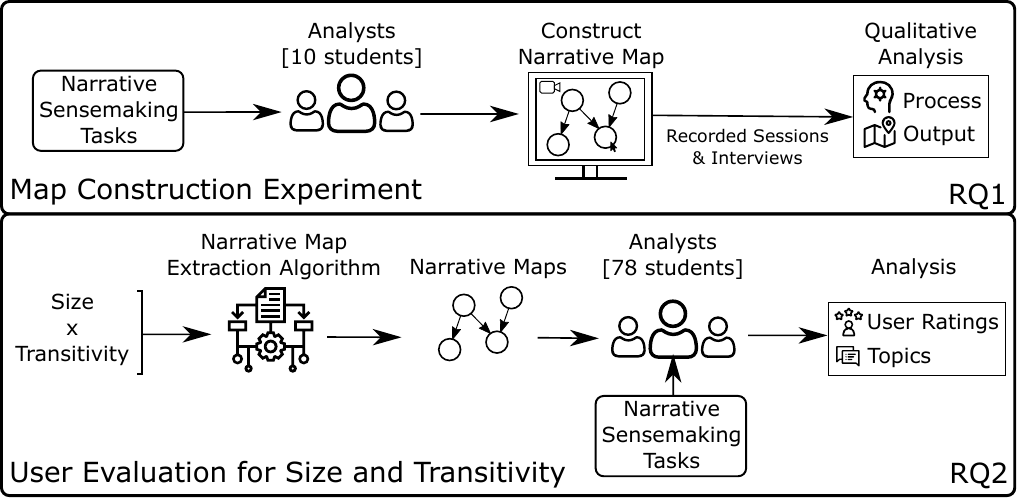}
    \caption{Overview of the experiments. The map construction experiment was used to answer RQ1. The user evaluation for size and transitivity was used to answer RQ2.}
    \label{fig:rqs}
\end{figure}

Figure \ref{fig:rqs} shows an overview of the experiments and research questions, which provides an overview of our experiments. We note that each of these research questions is also associated with a different type of user of narrative maps, while RQ1 is focused on users who create the maps, RQ2 is focused on users who only consume the maps without creating them. These users might have different needs, for example, map creators might want tools that make it easier to find new connections, and map consumers might prefer having additional interactivity to navigate the map. However, in both use cases, the narrative maps aid analysts in the connecting the dots task.

\section{Related Work}
\label{sec:rel-work}
First, we note that this work is an extended version of a short paper in a visualization conference \cite{norambuena2021narrative}. The original version included partial results and a more superficial analysis of our results for RQ1, focusing on connections types, construction strategies, and graph and layout properties. This extended version includes new insights on RQ1, such as event selection and additional features and suggestions proposed by the analysts. Furthermore, this version includes RQ2, which did not exist in the original publication. Finally, this version also includes a series of design guidelines for narrative maps and an in-depth discussion of our results.

In the rest of this section, we discuss the existing literature in the field of narrative visualization. In particular, we give a brief introduction to the intersection of narratives and visualization. Then, we discuss narrative extraction and representation methods. Finally, we discuss works that model cognitive strategies in the sensemaking process. 

\subsection{Narratives and Visualization}
Narratives are systems of stories interrelated with coherent themes \cite{halverson2011master}. These stories can be told in different ways, leading to a distinction between the story itself and how it is represented. Narrative studies attempt to understand the relationships between the underlying stories and their representations \cite{abbott2008cambridge, puckett2016narrative}. In the context of information visualization, we explore how information narratives and storylines can be visualized. 

Storytelling and narratives are common metaphors in visual analytics \cite{segel2010narrative,tong2018storytelling,riche2018data}. In general, scholars have studied how arranging visualizations as story sequences can be used to aid sensemaking \cite{hullman2013deeper,hullman2017finding}. Other works focused on narrative visualization for news usually focus on augmenting data visualization techniques (e.g., charts) with contextual information (e.g., relevant articles associated with data points in the chart) \cite{hullman2013contextifier, gao2014newsviews}. However, in our application context, we are interested in extracting and representing narratives taken directly from data sets of text documents, rather than augmenting numerical (or other non-text types of data) visualizations with contextual information or using sequences of visualizations to represent a story. Thus, not all of the visual storytelling concepts apply to our work, as they are designed for other types of visualizations in mind. Nevertheless, the visual storytelling framework provides a series of useful definitions \cite{segel2010narrative} as well as techniques and design patterns \cite{riche2018data} that could prove useful towards our goal of designing better narrative maps.

There are multiple genres of narrative visualizations. Narrative maps---and other graph-based narrative structures---provide paths that the users can follow to understand the story, similar to how flow charts work. Thus, they fall into the flow chart genre of narrative visualization, as defined by Segel et al. \cite{segel2010narrative}. Next, we consider the concept of messaging \cite{segel2010narrative} in visual storytelling, which refers to the use of text to provide explanations and observations about the visualization. In terms of messaging, narrative maps make heavy use of text, as the events in the maps are described entirely by text (e.g., the article's headlines) and annotations can be used to provide additional context for each part of the map. Finally, we note that interactivity \cite{segel2010narrative,tong2018storytelling} is another important element of visual storytelling, however, for the purposes of this paper, we did not consider interactive narrative maps in the evaluation. The study of interactivity in the context of narrative maps is left as future work.

Narrative maps usually show multiple storylines that can be visualized at the same time. Therefore, according to the storytelling taxonomy of Tong et al. \cite{tong2018storytelling}, narrative maps fall between the \textit{narrative visualization for storytelling in parallel} category or the \textit{narrative visualization overview} category. In this context, storytelling systems enable users to detect patterns, structures, or relationships in data, which can help users confirm hypotheses or gain additional knowledge about a specific topic \cite{akaishi2007narrative,tong2018storytelling}. We note that it would be possible to construct a map as a single timeline, leading to linear storytelling. However, this would be a pathological case and not the typical use case of narrative maps.

\subsection{Narrative Extraction}
Regardless of the underlying structure or representation used to model narratives, narrative extraction algorithms usually rely on optimizing different criteria, such as topical cohesion (whether connected events focus on the same topic) \cite{wang2015socially}, coherence (how much sense it makes to join two events) \cite{shahaf2010connecting}, or coverage (the proportion of the events covered by the narrative) \cite{shahaf2012trains}. In this work, we use a narrative extraction algorithm based on the criteria of coherence maximization through linear programming \cite{keith2020maps}. However, none of these narrative extraction algorithms are backed by an evaluation of how analysts construct narratives from data. Thus, in order to create better extraction algorithms, we seek to understand the narrative sensemaking process of analysts.

\subsection{Narrative Representation}
The core element of any narrative representation is an event, which is the basic unit of narratives as all stories are simply sequences of events in their most basic form \cite{abbott2008cambridge}. However, while most narrative representations focus on the event level \cite{shahaf2010connecting, liu2017growing, keith2020maps}, other representations do exist. One approach is to represent narratives in terms of topics, i.e., abstracting the narrative representation away from particular events and instead focusing on the overarching topics and how they relate to one another \cite{nallapati2004event,kim2011topic,zhou2017survey}. Some scholars have proposed more fine-grained resolution levels as well, such as individual named entities \cite{faloutsos2004fast}, the claims and attributions found in a news article \cite{soni2014modeling}, and hybrid resolution methods that would allow changing between levels in an interactive way \cite{shahaf2013information}. For the purposes of this work, we decided to focus on the event level, as this representation has strong theoretical foundations in narratology \cite{abbott2008cambridge} and they are the backbone of any narrative \cite{baikadi2011towards,kelter2004representing}.

There are three general approaches to structure narrative representations: timelines \cite{wang2015socially, yan2011evolutionary, shahaf2010connecting, shahaf2012connecting}, trees \cite{ansah2019graph, liu2017growing}, and directed acyclic graphs (DAGs) \cite{yang2009discovering, zhou2017emmbtt, shahaf2012trains,shahaf2013information, keith2020maps}. Moreover, these structures can be composed of a single connected structure \cite{shahaf2012trains,keith2020maps} or a series of disjoint and parallel structures (e.g., story forests) \cite{xu2013summarizing, liu2017growing}. 

The underlying representation of the narrative guides the visual design. For example, timeline approaches visually present the resulting narrative in a linear fashion, and most do not require advanced visualization techniques. In contrast. Structured approaches using trees or DAGs, in contrast, need more complex visualizations, such as information metro maps \cite{shahaf2013metro} or story forests \cite{liu2017growing}. Moreover, the different structures present trade-offs in terms of expressive power and complexity. For example, DAGs allow us to show divergent and convergent substructures, while trees only allow us to show divergent substructures. However, we still do not have a systematic evaluation of these different underlying structures. Thus, our work seeks to bridge this gap by exploring which one of these structures performs better in the context of narrative sensemaking.

\subsection{Cognitive Strategies in the Sensemaking Process}
Previous research has explored how analysts make cognitive connections between documents in the context of intelligence analysis tasks. For example, Bradel et al. \cite{bradel2013analysts} studied how analysts structure information in the context of intelligence analysis tasks, where they found layouts based on linear structures with branching and web-like structures. Our study also shares similarities with the work of Robinson \cite{robinson2008collaborative}, which focuses on analyzing the strategies and organizational methods used during collaborative synthesis, with the purpose of proposing a series of design guidelines for collaborative sensemaking systems. 

Other similar work includes Andrews et al. \cite{andrews2010space,andrews2012analyst}, who explore the workspace organization used by analysts in large displays to arrange documents, where most strategies consisted of clustering, although some analysts used timelines. In addition, Wenskovitch and North \cite{wenskovitch2020examination} study how analysts perform grouping and dimensionality reduction, where strategies included divide-and-conquer, incremental layouts, and bottom-up construction. Our work follows a similar approach, but focusing exclusively on the use of narrative maps as a sensemaking tool, analyzing the different map construction strategies and the underlying graph structures generated during the process.

Previous studies have also found that analysts use strategies such as identifying co-occurrence relationships and aggregating common elements \cite{haider2017analysts}, using topical and temporal orderings for document clustering, and evaluating content overlap and similarity for document summarization \cite{endert2012clustering, camargo2013manual}. However, previous research has not focused on specific narrative sensemaking tasks. In narratives, there is an underlying temporal ordering as well as a focus on cause-effect relationships, which leads to a specific description of cognitive connections and construction strategies for narrative sensemaking.

Finally, prior works have shown that graph-based narrative representations \cite{shahaf2013metro,keith2020maps,liu2017growing} are useful as a sensemaking tool. Thus, with the purpose of improving such narrative representations and their associated extraction algorithms, we seek to understand how analysts create such models from scratch by analyzing the narrative mapping process and its strategies.

\section{RQ1: Narrative Map Construction Strategies}
\label{sec:rq1}
In this section, we focus on answering \textbf{RQ1}. We first describe our user study. Then, we explain the event selection criteria used by the participants. After that, we study the types of cognitive connections used to construct the map. Next, we describe the different map construction strategies. Furthermore, we describe the graph and layout properties of the narrative map. Finally, we list some of the suggestions and features of narrative maps discussed by the participants. Throughout this process, we used \textit{open coding} \cite{khandkar2009open} to discover the different types of cognitive strategies and analyze the results. 

\subsection{Study Description}
\subsubsection{Data Set}
We used a data set comprised of 40 COVID-19 news articles from January 2020 that cover the start of the Coronavirus outbreak in all our experiments. This data set is a subset of the COVID-19 archive data used in previous works on narrative maps \cite{keith2020maps}. The events were carefully curated in order to have a sufficiently small data set for our manual map construction experiment while covering a series of different topics and issues regarding the COVID-19 narrative. In particular, the articles cover topics such as the economic consequences of the pandemic, the sociopolitical effects in China, the worldwide response, and others.
As our data set was made up of breaking news, the main event is usually described explicitly in the headlines \cite{norambuenaevaluating}. Thus, we focused on the headlines rather than the full article. We also included the publication dates and sources.

\subsubsection{Task Definitions}
\begin{figure*}[!htb]
    \centering
    \includegraphics[width=\textwidth]{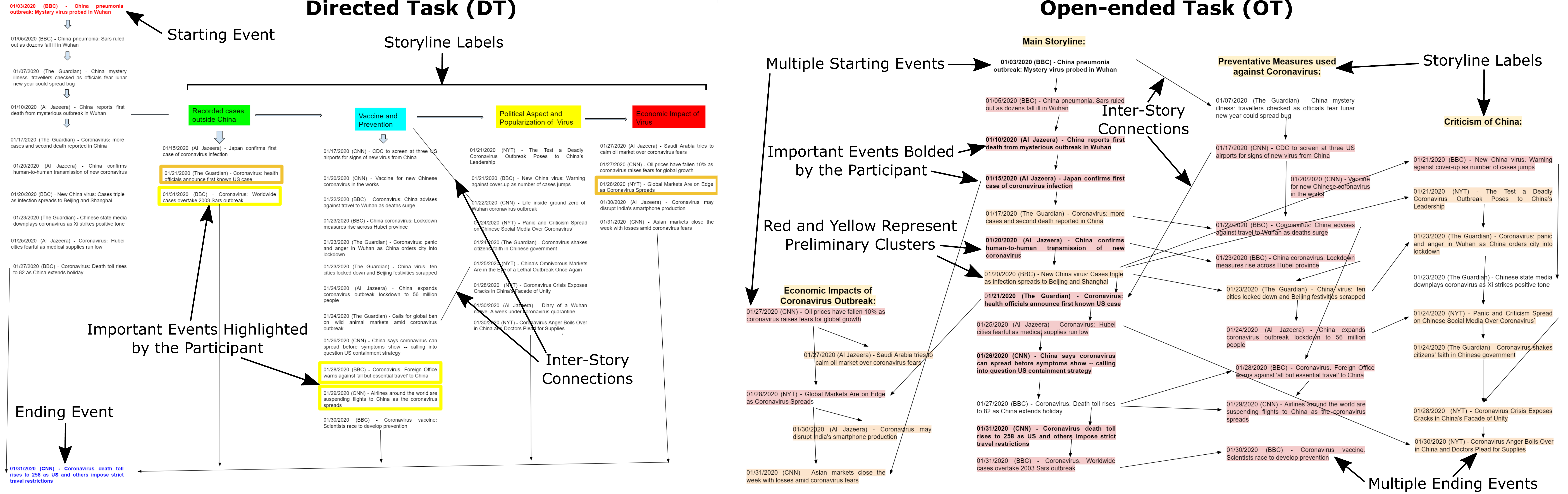}
     \caption{Narrative map examples created by participants for the two tasks \textbf{Directed Task} (DT) and \textbf{Open-ended Task} (OT). Annotations highlight key elements in these maps. Note how in the DT map all storylines converge into a single ending event, emphasizing the focused nature of this task. In contrast, the OT map has a series of storylines that interact with one another, representing the different outcomes found by the participant.}
    \label{fig:examples-rq1}
\end{figure*}

As in our motivational example, we defined two tasks to explore how analysts constructed narrative maps, a directed task that required participants to join two events and an open-ended task that required participants to expand on the outcomes of an initial event (see examples in Figure \ref{fig:examples-rq1}). In both tasks, participants were given a list of events (i.e., nodes) and asked to construct a narrative map by designing its overall structure, layout, and specific connections. The participants were also asked to label their main storyline---the core events of the narrative---and their side stories---stories relevant to the overall narrative but not directly related to the main storyline. The focus of this experiment was to glean insights on the construction process, rather than comparing how the tasks themselves influence the construction. By considering two tasks rather than a single one, we expected to gather additional insights regarding the construction of narrative maps.

The directed task required participants to construct a narrative map to answer the following question: ''\textit{How did the Wuhan outbreak lead to the US travel restrictions?}", which referred to two specific events in the data set. This task is also known as ``connecting the dots" and it is a fundamental task in narrative sensemaking \cite{keith2020maps}. Previous research has attempted to understand how analysts perform this process \cite{bradel2013analysts} and sought to automate this process through algorithmic approaches \cite{shahaf2010connecting}. Note that while users are allowed to create side stories, the focus is on finding the connections between the two events rather than on finding other outcomes.

In contrast, the open-ended required participants to construct a narrative map to answer the following question: ''\textit{What outcomes occurred as a result of the Wuhan outbreak?}". This task is a variation of the basic ``connect the dots" task \cite{shahaf2012connecting} that only provides the starting event as a fixed point, requiring the participants to explore the storylines that emerge because of this event. The focus is on finding storylines and outcomes in the narrative, rather than connecting two specific events. We designed this task to allow participants more degrees of freedom in their analysis, letting them define what they consider to be an important outcome. More specifically, 

Both tasks required participants to label their storylines and to answer a follow-up question with their map: ``\textit{What are the key events (i.e., the most important events or turning points)?}". All other instructions and examples were the same for both tasks. The only difference being the basic question that guides the map construction process. 

Finally, we note that the tasks defined for this experiment represent simplified and constrained versions of what analysts would do in a real-world setup. In particular, they exclude the foraging loop from the sensemaking process, as we provide a pre-selected and curated data set. Moreover, they all use the same document as a starting point. These constraints were imposed in order to the make analysis simpler by eliminating the additional complexity and variables that foraging and unguided analysis could create. Thus, the created maps are easier to compare and analyze. Regardless of these constraints, the tasks still provide valuable insights into narrative sensemaking, and more specifically into the synthesis loop of the sensemaking process.

\subsubsection{Evaluation Procedure}
We recruited ten participants, following a similar approach to the work of Bradel et al. \cite{bradel2013analysts}. We assigned five participants to each task. While splitting the participants into two tasks increases variability, we expected to gather a wider range of construction strategies by doing this. All participants were advanced undergraduate students part of a national security program and hence, had a background in intelligence analysis. They also had previous knowledge on the topic which they could leverage while conducting the tasks. Prior knowledge ranged from general knowledge about COVID-19 to stronger backgrounds since some participants were ardent followers of the pandemic news right from its start. Figure \ref{fig:examples-rq1} shows examples of the maps created in each task.

To provide initial training and to avoid inducing biases in subsequent task performance, participants were provided with a short example narrative map on a different topic. We engaged with our participants in an hour-long semi-structured session in a video call where they completed their assigned narrative sensemaking task. Participants were provided with a short example of a narrative map to guide them. The example map was on a different topic to avoid inducing biases in potential connections. We encouraged the participants to think aloud and ask questions and share any observations as they worked. We explained that there were no correct or incorrect answers; as our goal was to understand the cognitive strategies used by the analysts to complete the tasks. However, the quality and conceptual cohesion varied among the solutions. All participants were recorded and the videos were analyzed to understand their construction strategies. In particular, we used \textit{open coding} \cite{khandkar2009open} to perform a qualitative analysis of the created maps and the sessions themselves.

To construct the map, we gave participants a canvas on Google Drawings with the instructions and the list of articles chronologically ordered. The participants had to drag and drop the articles into the available space. Then, they had to add connections with arrows. The participants were instructed to design the map with other analysts as potential users in mind. The participants were familiar with Google Drawings and similar editing tools, thus they did not require additional training in its use, even if it might not have been their preferred tool for such an exercise. Moreover, they had full access to this tool through their institutional accounts. 

We opted for Google Drawings in our study for several reasons. First, it provided a closer approximation of what a computational narrative map tool would look like compared to an approach using hand-drawn notes. Thus, even though it might influence the kinds of strategies used by the participants, these strategies should be closer to what we would expect with a computational tool. Second, given the limitations caused by the pandemic, using Google Drawings allowed us to do virtual sessions, thus minimizing the risks for the participants. Finally, it also provided a detailed editing history which, in conjunction with the recorded sessions, was useful to precisely analyze the steps taken by the participants.

\subsection{How do analysts select events?}
We asked participants to explain their event selection process during the creation of the map. Table \ref{tab:event-sources} shows the results for each analyst.

\begin{table}[!htb]
\centering
\resizebox{\columnwidth}{!}{%
\begin{tabular}{@{}cccccccccccc@{}}
\toprule
\textbf{Property}                       & \textbf{Code}     & \textbf{D1} & \textbf{D2} & \textbf{D3} & \textbf{D4} & \textbf{D5} & \textbf{O1} & \textbf{O2} & \textbf{O3} & \textbf{O4} & \textbf{O5} \\ \midrule
\multirow{3}{*}{\textbf{Important Events}} & Hard Facts        & $\times$    &             &             &             &             &             & $\times$    & $\times$    & $\times$    &             \\
                                        & Perceived Impacts &             & $\times$    & $\times$    &             & $\times$    &             &             &             &             & $\times$    \\
                                        & Map Structure     &             &             &             & $\times$    & $\times$    & $\times$    &             &             &             &             \\ \bottomrule
\end{tabular}%
}
\caption{Selection criteria for important events.}
\label{tab:event-sources}
\end{table}


First, regarding the selection of important events, participants either focused on ``hard facts'' (e.g., number of deaths and scientific information), the ``perceived impacts'' of an event (e.g., panic and social issues), or the map structure (e.g., number of connections or how an event summarizes the surrounding events). Four participants focused on hard facts and avoided referring to opinions or speculations in their selection of important events, as they wanted their narrative to be as objective as possible. This included reporting events such as the number of deaths, statistics, scientific information, and government responses. In contrast, the four participants who focused on ``impacts'' did not shy away from opinion-based or speculative headlines, since these events might provide insight into the actual perceived impacts of the outbreak. The directed-task participants that focused on impacts explicitly mentioned that they were concerned with the impact concerning the travel restrictions, as the directed task made them focus on this issue. The open-ended participant that focused on impacts used their own experiences with the virus to determine impacts. Finally, those who focused on the map structure selected the events based on their context in the underlying graph, considering whether the event acted as a hub node or whether it served as a summary of its surrounding articles or storyline.

Next, we explored how participants used the information regarding the news source of each event during the event selection process. Most participants did not use the sources, with some of them outright ignoring them. Reasons vary from ``the sources are filtered and reliable enough" to simply ``I was focused on the dates and headlines". Most participants found that the sources were reliable enough and as they were relatively mainstream sources, they did not question their content. In this context, some participants commented that a narrative map should have more sources and that the sources should be balanced to prevent introducing biases in the narrative (e.g., having all sources come from one side of the political spectrum). In particular, some participants suggested limiting the sources to mainstream media.

The actual usage of the news source information varies. For example, a participant used his knowledge about the BBC to determine that one of the articles referred to a governmental office in the UK. Someone found the early Al Jazeera articles on the virus as an important sign indicating the spread and impact of the virus. Thus, we found that the news sources did not influence the selection of events or their connections, at least with this data set. 

We note that in a real-world application the quality of the sources would be a very important consideration for analysts, which might affect the results of such experiments. However, in this experiment, the data was pre-selected and curated, as our work did not focus on the foraging loop of the sensemaking process. Instead, our goal was to understand the narrative structures that analysts would create, rather than how they would filter and collect the data and sources. Thus, for the purpose of this experiment, only mainstream and reputable sources were selected, in order to avoid the additional layer of complexity of dealing with biased or untrustworthy news sources.

\begin{table}[!htb]
\centering
\resizebox{\columnwidth}{!}{%
\begin{tabular}{@{}C{6cm}ccc@{}}
\toprule
\textbf{Events}                                                                                             & \textbf{Directed Task} & \textbf{Open-Ended Task} & \textbf{Global} \\ \midrule
China pneumonia: Sars ruled   out as dozens fall ill in Wuhan                                               & 0.4                    & 0.6                      & 0.5             \\
China reports first death from mysterious outbreak in Wuhan                                               & 0.8                    & 1                        & 0.9             \\
Japan confirms first case of coronavirus infection                                                        & 0.2                    & 0.6                      & 0.4             \\
Coronavirus: more cases and second death reported in China                                                & 0.8                    & 0.8                      & 0.8             \\
CDC to screen at three US airports for signs of new virus from China                                      & 0.6                    & 0.2                      & 0.4             \\
Vaccine for new Chinese coronavirus in the works                                                          & 0.2                    & 0.6                      & 0.4             \\
China confirms human-to-human transmission of new coronavirus                                             & 1                      & 1                        & 1               \\
New China virus: Cases triple as infection spreads to Beijing and Shanghai                                & 0.8                    & 0.8                      & 0.8             \\
Coronavirus: health officials announce first known US case                                                & 0.8                    & 0.6                      & 0.7             \\
China virus: ten cities locked down and Beijing festivities scrapped                                      & 0.2                    & 0.6                      & 0.4             \\
China says coronavirus can spread before symptoms show -- calling into question US containment strategy & 0.6                    & 0.6                      & 0.6             \\
Coronavirus: Death toll rises to 82 as China extends holiday                                              & 0.4                    & 0.6                      & 0.5             \\
Coronavirus: Worldwide cases overtake 2003 Sars outbreak                                                  & 0.8                    & 0.6                      & 0.7             \\
\midrule
Total events with at least 0.5 frequency   & 8 & 12 & 7\\
\bottomrule
\end{tabular}%
}
\caption{Events that were selected by a majority of the participants in at least one task. The first column shows the event, the second and third column show the frequency for that particular task, and the last column shows the global frequency. Note that only one event is common to both tasks (human-to-human transmission event).}
\label{tab:frequencies}
\end{table}

We then turned our attention towards the events that were selected by the participants. We present the most common ones in Table \ref{tab:frequencies} (i.e., those selected by a majority of participants in at least one task). The directed task had fewer common events than the open-ended task, which could be due to the nature of the directed task requiring to focus specifically on how the outbreak led to the US travel restrictions. However, the event regarding human-to-human transmission was considered by all participants in their narrative map.

Finally, we studied the alignment between participants in terms of included and excluded events. For each event, we measured the number of times they were included and excluded in the maps. Then, we took the maximum value among these two and averaged over all events. This gave us the average alignment among all participants. The best possible value of alignment would be 1.0, which means that either all participants agreed that it should be included or excluded. The worst possible value of alignment would be 0.5, as that would mean that the event is equally included and excluded by the participants. Following this approach, we find that the directed task has an alignment of 76.32\% (excluding the pre-defined starting and ending events). In contrast, the open-ended task only has an alignment of 56.41\% and much higher variability in terms of event inclusion and exclusion. This makes sense as the directed task gives a specific guiding question to the participants.

\subsection{How do analysts connect events?}
To answer this question, we asked participants to explain their connection strategies as they constructed the map as well as in the follow-up interviews. We identified seven types of connections, which we further divided into low-level, high-level, and supporting connections. Low-level connections are those that can be made directly from the content of the document (e.g., dates, keywords, entities present) without an in-depth analysis. In contrast, high-level connections involve applying cognitive schemas to synthesize information between events \cite{bradel2013analysts}. Supporting connections are used in conjunction with high-level connections as an auxiliary strategy to help connect events. For example, a connection could be based on cause-effect relationships between events (high-level connection) and analyst speculation (supporting connection). Table \ref{tab:connections} summarizes the different connection types and Table \ref{tab:connections-results} shows the results for the different connection types for each analyst.

\begin{table*}[!htb]
\centering
\resizebox{0.75\textwidth}{!}{%
\begin{tabular}{@{}cccc@{}}
\toprule
\textbf{Type}                          & \textbf{Code}   & \textbf{Description} & \textbf{Number of Maps} \\ \midrule
{\textbf{Low-level}}  & Temporal & $A$ happened before $B$ & 10             \\
                                        & Similarity & $A$ is similar to $B$ & 7              \\
                                        & Entity & $A$ is about the same entity as $B$ & 4              \\ \midrule
{\textbf{High-level}} & Topical & $A$ shares a common theme or topic with $B$ & 10             \\
                                        & Causal & $A$ leads to $B$ & 7              \\ \midrule
{\textbf{Supporting}}
                                        & Domain Knowledge & $A$ is related to $B$ because of external knowledge $X$ & 4              \\
                                        & Speculative & $A$ is connected to $B$ because of inferred $X$ & 3              \\ \bottomrule
\end{tabular}%
}
\caption{Connection types for each participant in our user study.}
\label{tab:connections}
\end{table*}

\begin{table}[!htb]
\centering
\resizebox{\columnwidth}{!}{%
\begin{tabular}{@{}ccccccccccccC{5cm}@{}}
\toprule
\textbf{Type}        & \textbf{Code}         & \textbf{D1} & \textbf{D2} & \textbf{D3} & \textbf{D4} & \textbf{D5} & \textbf{O1} & \textbf{O2} & \textbf{O3} & \textbf{O4} & \textbf{O5}\\ \midrule
{\textbf{{Low-level}}}     & Temporal         & $\times$  & $\times$  & $\times$  & $\times$  & $\times$  & $\times$  & $\times$  & $\times$  & $\times$  & $\times$   \\
                         & Similarity       & $\times$  & $\times$  & $\times$  & $\times$  &    & $\times$  &    & $\times$  & $\times$  &   \\
                         & Entity           & $\times$  & $\times$  &    & $\times$  &    & $\times$  &    &    &    &    \\ \midrule
{\textbf{{High-level}}}    & Topical          & $\times$  & $\times$  & $\times$  & $\times$  & $\times$  & $\times$  & $\times$  & $\times$  & $\times$  & $\times$  \\
                         & Causal           & $\times$  &    & $\times$  & $\times$  &    & $\times$  &    & $\times$  & $\times$  &  $\times$  \\\midrule 
{\textbf{{Supporting}}}
                         & Domain Knowledge &    &    &    &    & $\times$  &    & $\times$  &    & $\times$  &  $\times$  \\ 
                         & Speculative      &    &    & $\times$  &    &    &    &    & $\times$  & $\times$  &  \\ \bottomrule
\end{tabular}%
}
\caption{Connection types for each participant in our user study.}
\label{tab:connections-results}
\vspace{-15pt}
\end{table}

\subsubsection{Low-level Connections}
We identified 3 low-level connections: temporal, similarity, and entity.

\begin{enumerate}[wide, labelindent=0pt]

    \item \textbf{Temporal connections} (A happened before B): Most participants used temporal connections as their default connection strategy. 
    In particular, this type of connection was used when there were no other explicit relationships between events and participants wanted to maintain the temporal sequence of the events (e.g., ``I just followed chronological order"). All participants used temporal connections in their maps, because of the inherent chronology of narratives \cite{baikadi2012towards}. However, some included event connections that broke the explicit chronological order. These non-chronological connections were a result of inferred causality relations---discussed later in the high-level connections. 
    
    \item \textbf{Similarity connections} (A is similar to B): 
    Users determined two events as similar primarily based on keyword matching and a superficial similarity evaluation (e.g., ``All these events mentioned markets or production''). These connections can be seen as a low-level counterpart of the topical connections discussed later. In computational terms, these connections can be easily implemented through text similarity functions or keyword matching techniques. Moreover, we note a special case of similarity connections, where documents that explicitly refer to the same event or that are too similar are grouped together as a single document (e.g., ``Calls for a global ban on wild animal markets amid coronavirus outbreak" and ``China's Omnivorous Markets Are in the Eye of a Lethal Outbreak Once Again'' could refer to the same event). 
    
    \item \textbf{Entity connections} (A is about the same entity as B): These connections are based on named entity co-occurrences in events. For example, some participants focused on whether the events referred to specific entities; in this case, the entities mostly referred to the countries being affected (e.g., ``These events talk about China"). Such connections could be extracted computationally by using named entity recognition techniques. We note that entity-based connections have been identified before as one of the common low-level techniques that analysts use to ``connect the dots" between documents \cite{bradel2013analysts}. 

\end{enumerate}

\subsubsection{High-level Connections}
We classified connections as high-level if they involved the use of a cognitive schema to connect information between documents. In particular, these connections arise usually from inferences made by the users rather than a superficial characteristic of the document. We identified two high-level connections in the user-generated maps. 

\begin{enumerate}[wide, labelindent=0pt]
\item \textbf{Topical connections} (A shares a common theme or topic with B): These connections are a more abstract version of the similarity connection. They focus on the overarching topic or theme of the articles (e.g., ``These events are about the Chinese government response"). These connections are distinct from the low-level similarity ones because they are based on a semantic viewpoint rather than superficial similarity. From a computational perspective, topical connections can be implemented through a topic modeling or clustering approach. In particular, we could determine whether the events fall under the same topic or cluster. A special type of topical connection is based on how information is presented, rather than on the topic itself. For example, events that share a specific media frame \cite{de2005news} could be related by this connection (e.g., ``These headlines criticize the Chinese government"). Based on the idea of media frames, we call this subtype of connection a framing connection. Moreover, follow-up questions suggest that this connection might be more useful for specific kinds of articles (e.g., op-eds). 

\item \textbf{Causal connections} (A leads to B): These are high-level relationships that join events if one is caused (or could be caused) by another (e.g., ``The number of cases surpassing SARS led to stricter travel restrictions"). 
Causal connections also cover events that could be reactions to another event, even if they are not explicitly caused by it. 
A special type of causal connection used by one of the participants corresponds to a ``supporting argument" connection that joins events if one of them provides a supporting argument for the conclusion or occurrence of another event, even if they do not directly cause it (e.g., ``The foreign office warning against travel supports the decision of airlines to suspend flights"). Finally, some causal relationships might defy temporal ordering because the reporting date of an event is not the same as the date when the event happened. In those cases, the participants changed the order of the events to respect the cause-effect relation.
\end{enumerate}

\subsubsection{Supporting Connections}
We classified connections as supporting if they are auxiliary strategies used in conjunction with a high-level connection. In particular, we identified two supporting connections in the user-generated maps.

\begin{enumerate}[wide, labelindent=0pt]
\item \textbf{Speculative connections} (A is connected to B because of inferred X): These connections are based on implications based on the participant's beliefs (e.g., ``The call for a global ban on animal markets made the global markets panic", a paraphrased causal speculative connection from one of the participants). Thus, speculative connections relate events that do not share any explicit relationship but could be connected based on speculative reasons. Note that these reasons might be right or wrong, but what matters is their speculative nature. 

\item \textbf{Domain Knowledge connections} (A is related to B because of external knowledge X): These connections are a special type of connection where documents that do not share any explicit relationship are connected based on external domain knowledge (e.g., ``Air travel and oil demand are related"). Note that these reasons might be right or wrong, but what matters is the dependence on external knowledge. 
\end{enumerate}

\subsection{What are analysts' map construction strategies?}
\begin{figure}[!htb]
    \centering
    \includegraphics[width=\columnwidth]{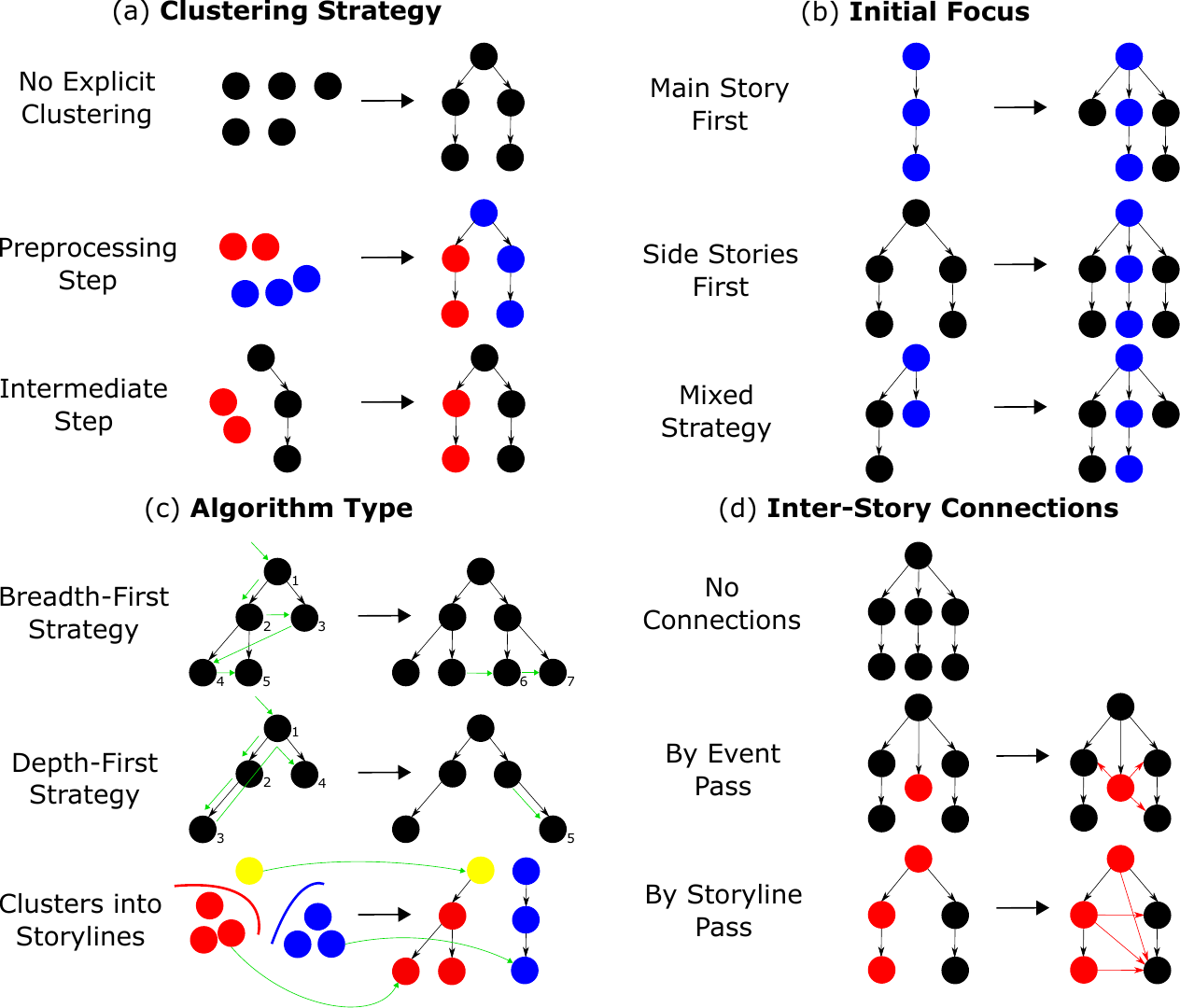}
    \caption{Narrative map construction strategies. (a) \textbf{Clustering strategies}: no clustering, clustering during preprocessing, or clustering in the middle of construction. b) \textbf{Initial focus}: whether participants created the main story (in blue) first, the side stories first, or a mixed strategy. (c) \textbf{Algorithm type}: the order in which nodes were added, following either a breadth-first, depth-first, or clustering approach. (d) \textbf{Inter-story connections}: some participants checked for inter-story connections (in red) when adding events, others checked when completing a storyline.}
    \label{fig:strats}
\end{figure}

We studied the construction process by following the individual steps taken by the participants as they built their narrative maps. We also asked follow-up questions about the process during the interviews. We identified a series of construction strategies for each analyst, that we display in Table \ref{tab:strat}. We also display a diagrammatic overview of the different strategies in Figure \ref{fig:strats}. Note that these strategies are abstract versions of the actual strategies that were obtained after analyzing the narrative map construction process step by step. Thus, these models provide a general idea of the construction strategy followed by participants, although there might be minor differences in some steps.

\begin{table*}[!htb]
\centering
\resizebox{0.75\textwidth}{!}{%
\begin{tabular}{@{}cccccccccccc@{}}
\toprule
\textbf{Property}                                 & \textbf{Code}                     & \textbf{D1} & \textbf{D2} & \textbf{D3} & \textbf{D4} & \textbf{D5} & \textbf{O1} & \textbf{O2} & \textbf{O3} & \textbf{O4} & \textbf{O5} \\ \midrule
\multirow{3}{*}{\textbf{\textit{Clustering Strategy}}}     & No Explicit Clustering   &    & $\times$  &    &    & $\times$  &    & $\times$  &    & $\times$  &  $\times$  \\
                                         & Preprocessing Step       &    &    & $\times$  & $\times$  &    &    &    & $\times$  &    &    \\
                                         & Intermediate Step        & $\times$  &    &    &    &    & $\times$  &    &    &    &    \\ \midrule
\multirow{3}{*}{\textbf{\textit{Initial Focus}}}         & Main Story First         & $\times$  &   &    &    & $\times$  & $\times$  & $\times$  &    &   &    \\
                                         & Side Stories First       &    &    & $\times$  &    &    &    &    &    &    &  $\times$  \\
                                         & Mixed Strategy           &    & $\times$   &    & $\times$  &    &    &    & $\times$  & $\times$   &    \\ \midrule
\multirow{3}{*}{\textbf{\textit{Algorithm Type}}}          & Breadth-First Strategy   &    & $\times$  &    &    &    &    &    &    & $\times$  &    \\
                                         & Depth-First Strategy     & $\times$  &    &    &    & $\times$  & $\times$  & $\times$  &    &    & $\times$   \\
                                         & Clusters into Storylines &    &    & $\times$  & $\times$  &    &    &    & $\times$  &    &    \\ \midrule
\multirow{3}{*}{\textbf{\textit{Inter-story Connections}}} & No Connections           &    &    & $\times$  & $\times$  & $\times$  & $\times$  & $\times$  &    &    &  $\times$  \\
                                         & By Event Pass            &    & $\times$  &    &    &    &    &    &    & $\times$  &    \\
                                         & By Storyline Pass        & $\times$  &    &    &    &    &    &    & $\times$  &    &    \\ \bottomrule
\end{tabular}%
}
\caption{Construction strategies for each participant in our user study.}
\label{tab:strat}
\end{table*}

\begin{enumerate}[wide, labelindent=0pt]

\item \textbf{Clustering Strategy}: Clustering allows analysts to group documents based on specific characteristics (e.g., topic, type of document, source). Half of the participants had an explicit clustering step during the creation of the map. The use of clustering in sensemaking tasks has also been reported in previous research, either as a story construction strategy \cite{bradel2013analysts} or as the final product \cite{endert2012clustering}. However, in the context of narrative maps, the main purpose of clustering is as a tool to aid in storyline constructions, without explicitly appearing in the final narrative map in most cases.
Clustering was either done as a preprocessing step (i.e., before starting with the connections) or as an intermediate step (i.e., after starting with the connections) to help identify storylines. Not all documents were clustered by people, although one of the participants did cluster all documents before creating the map. See Figure \ref{fig:clusters} for examples of both clustering strategies. 

\begin{figure}[!htb]
    \centering
    \begin{minipage}[b]{0.5\textwidth}
        \centering
        \includegraphics[width=\textwidth]{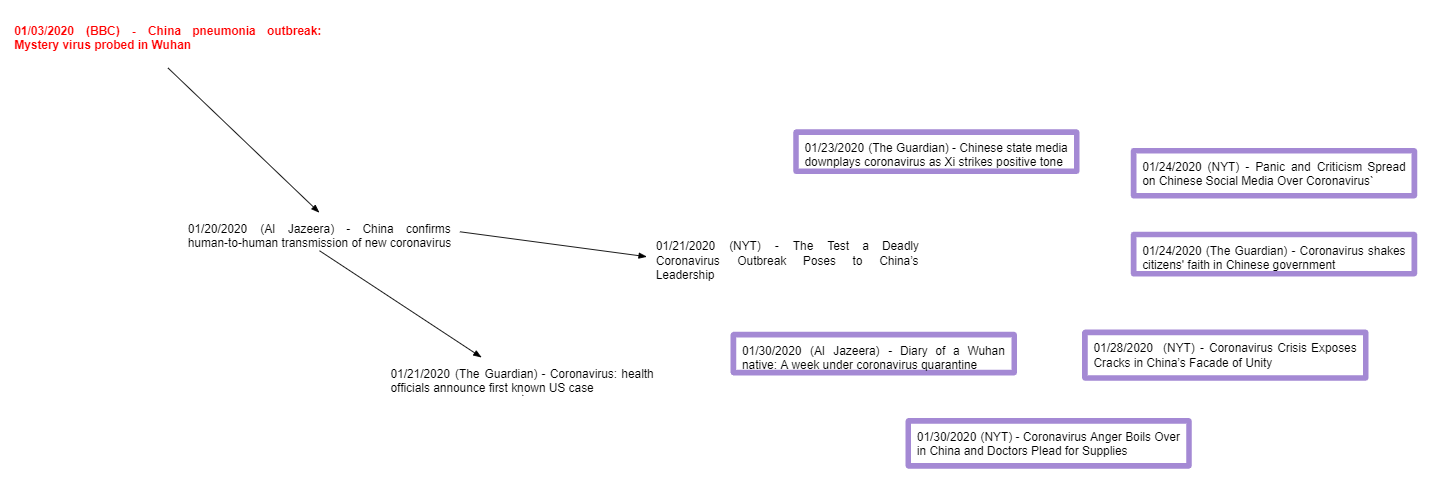}
    \end{minipage}\\%
    \begin{minipage}[b]{0.5\textwidth}  
        \centering 
        \includegraphics[width=\textwidth]{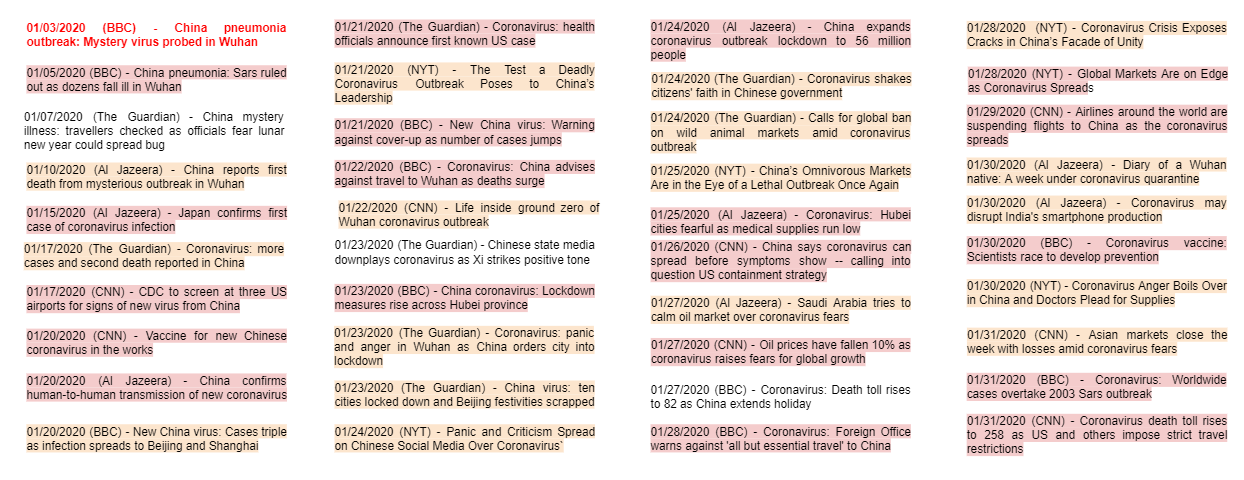}
    \end{minipage}%
    \caption{Examples of different clustering strategies. The top example shows the creation of a cluster through an \textbf{intermediate clustering step} during map construction (the highlighted events are about "Chinese Media Criticism"). The bottom example shows \textbf{clustering as preprocessing}, where cluster labels are assigned before constructing the map (the highlighted events were initially classified with respect to the questions asked in the instructions).}
    \label{fig:clusters}
\end{figure}

Moreover, some clusters changed over time (e.g., adding new documents to an existing cluster as the map was created) while others remained static (i.e., the clusters did not change after creation). Finally, there are also cases where the participants did not perform any explicit clustering step. For participants performing the directed tasks, there was a slight preference for clustering in comparison to the open-ended task participants. This could be due to the directed nature of the task, which could have allowed participants to define clusters more easily, as the guiding question could be answered by grouping events that focused on travel or the US. In contrast, the open-ended task did not provide any explicit guidelines for cluster formation.

\item \textbf{Initial Focus}: This strategy refers to the part of the narrative map that was created first. Participants either focused on the main story, the side stories, or followed no particular order (i.e., a mixed strategy going back and forth). The main story refers to the sequences of core events in the narrative, those that move the narrative forward \cite{abbott2008cambridge}. In contrast, the side stories do not form part of the narrative core. Instead, they provide further information and useful context to the narrative. Note that there is an even split between focusing on the main story and following a mixed strategy. Only two participants decided to focus on the side stories first. 

\item \textbf{Algorithm Type}: This strategy refers to the general algorithm that participants followed to construct the map. By analyzing the order in which participants constructed the maps, we found three types of strategies. The first two strategies are conceptually similar to basic graph searching algorithms---constructing the map in a depth-first or breadth-first fashion---while the third strategy is based on clustering---turning clusters into storylines. Note that depth-first approaches either focused on side stories first or on the main story first. In contrast, breadth-first approaches followed a mixed strategy by definition. The strategy of turning clusters into storylines either focused on side stories first or followed a mixed strategy. We note that it would be technically possible to focus on the main story when using the clusters into storylines approach. However, we did not observe this behavior during our experiments. 

\item \textbf{Inter-story Connections}: This strategy refers to how the participants connected storylines. In most cases, participants did not add inter-story connections; making their storylines independent from the rest of the graph, except for the initial connection where they split off. In other cases, they added connections on a by-event basis, checking whether an event should be connected to other stories as they add it. Alternatively, they added connections on a by-storyline basis, checking whether to connect the storyline with others only after completing the whole storyline. For example, the inter-story connections in the open-ended map of Figure \ref{fig:examples-rq1} were added after the storylines were completed. We note that, in general, there were few inter-story connections, as each storyline was clearly defined and mostly self-contained.
\end{enumerate}

\subsection{What are the properties of the created maps?}
We answer by focusing on multiple structural aspects of the underlying graph and the layout considerations made by participants (see Table \ref{tab:layout}). 

\begin{enumerate}[wide, labelindent=0pt]
\itemsep0em
\item \textbf{Graph Structure}: We found that participants used three types of underlying graph structures: lists \protect\resizebox{0.35cm}{!}{\protect\includegraphics{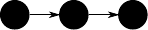}}, trees \protect\resizebox{0.35cm}{!}{\protect\includegraphics{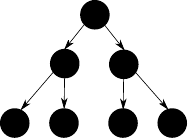}}, and directed acyclic graphs (DAGs) \protect\resizebox{0.25cm}{!}{\protect\includegraphics{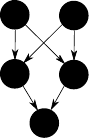}}. These results are in line with prior work on story and narrative representations, which has focused on similar types of structures to represent stories \cite{keith2020maps,shahaf2012trains}, such as timelines \cite{shahaf2010connecting}, trees \cite{ansah2019graph}, or other graph variants \cite{yang2009discovering}. For our study, structures were evenly split between trees and DAGs, with only two list-like graphs, where one of them was a single timeline and the other comprised three parallel timelines \protect\resizebox{0.35cm}{!}{\protect\includegraphics{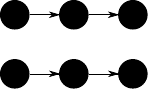}}. The person who used the single list structure explained that they were trying to create a timeline that covered the important events, rather than expanding on side stories.

\item \textbf{Layout and Main Story Position}: 
Most participants went for a vertical (top-down) approach \protect\resizebox{0.25cm}{!}{\protect\includegraphics{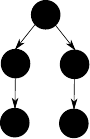}} with storylines presented as parallel columns and the main story placed first \protect\resizebox{0.30cm}{!}{\protect\includegraphics{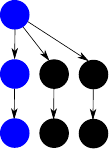}} (i.e., the left-most story in a vertical layout or the top story in a horizontal layout). Horizontal layouts \protect\resizebox{0.35cm}{!}{\protect\includegraphics{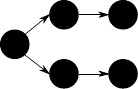}} were not preferred; as noted by participants, computer displays seem to favor vertical layouts due to how scrolling works. Finally, one participant used a unique diagonal layout \protect\resizebox{0.35cm}{!}{\protect\includegraphics{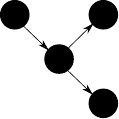}} (shown in Figure \ref{fig:diagonal}; we did not observe this behavior in any of the other participants.

\begin{figure}[!htb]
    \centering
    \includegraphics[width=\columnwidth]{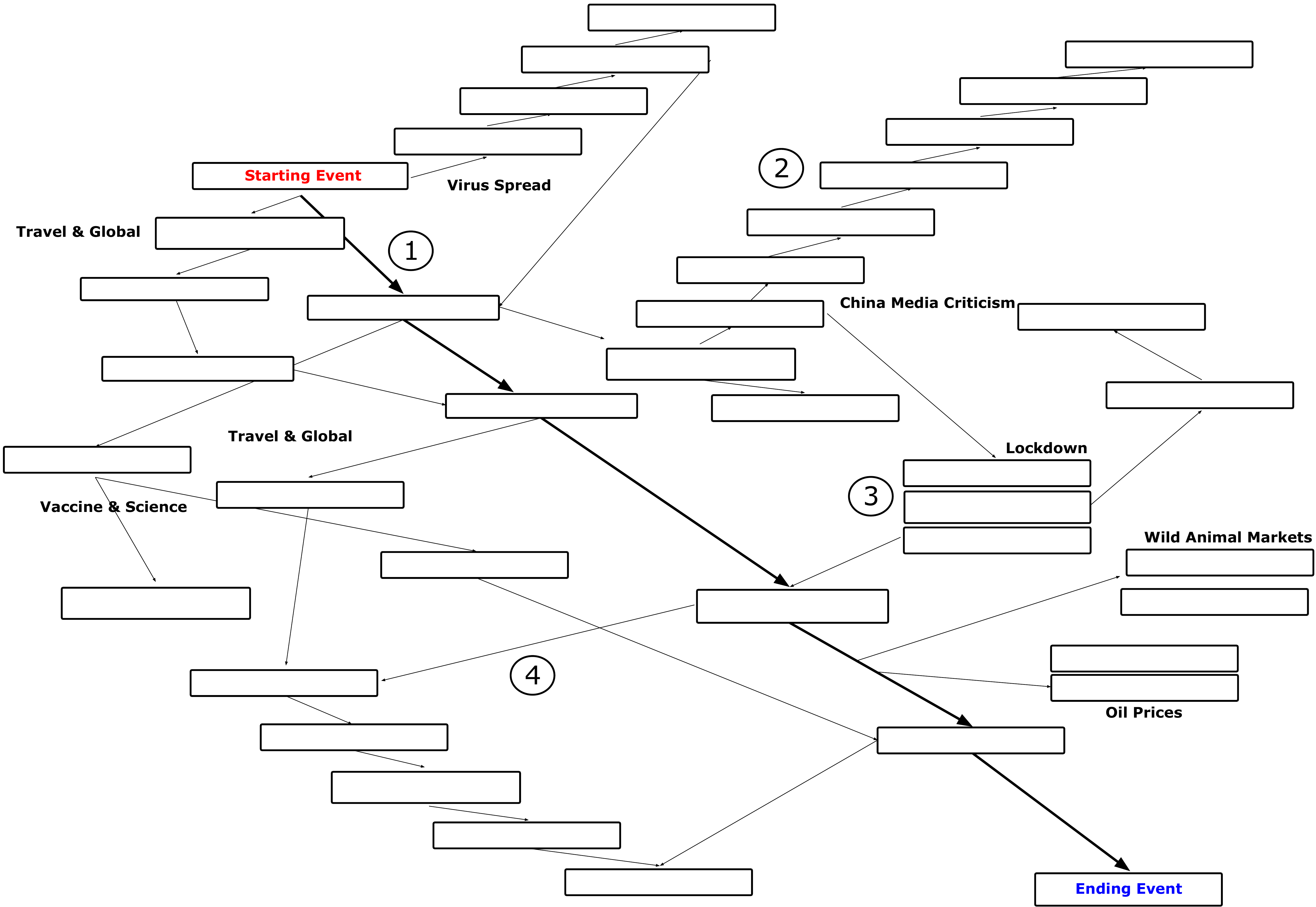}
    \caption{Example narrative map structure from a participant of the Directed Task. Note that the map has a diagonal layout---the only map that uses this type of layout---with its main story (1) on its center. Moreover, this map was constructed following a depth-first strategy, starting with the main story and then branching into the side stories (2). Some events that were considered too similar or the same were grouped together into a single block (3). Inter-story connections (4) were added following a by storyline pass strategy.}
    \label{fig:diagonal}
\end{figure}

\item \textbf{Number of Source and Sink Nodes}: Most people had multiple storyline endings (i.e., sink nodes) \protect\resizebox{0.25cm}{!}{\protect\includegraphics{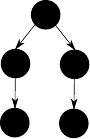}}. In particular, all five open-ended task maps had multiple endings. In contrast, the directed task had two participants constrain themselves to a single ending as defined by the task \protect\resizebox{0.25cm}{!}{\protect\includegraphics{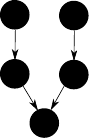}}, while the others added endings or dead-end events for some of the other storylines. For source nodes, participants that had the directed task were more likely to have a single source \protect\resizebox{0.25cm}{!}{\protect\includegraphics{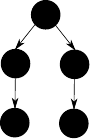}} than those that had the open-ended task \protect\resizebox{0.25cm}{!}{\protect\includegraphics{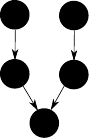}}. The tendency of open-ended maps to have multiple sources and sinks intuitively makes sense given the unrestricted nature of the task. In contrast, the directed task maps are naturally more focused on just answering the main question (``How did the Wuhan outbreak lead to the US travel restrictions?"), thus leading to structures that did not have as many loose ends.

\item \textbf{Connectivity}: Most participants created connected graphs \protect\resizebox{0.25cm}{!}{\protect\includegraphics{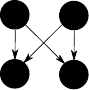}}. In graph-theoretical terms, we classify the map as connected if its underlying graph is weakly connected (i.e., we disregard the direction of the arrows). 
However, there were two cases where the graphs had separate components \protect\resizebox{0.25cm}{!}{\protect\includegraphics{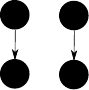}}. The first had a separate component for the ``social response and effects of COVID" that was not connected to any other story. The second had three parallel timeline structures (the main story, economic effects, and preventive measures) without any explicit connection between them. 

\item \textbf{Transitivity}: We considered whether participants explicitly included connections that are implied by transitivity (i.e., \protect\resizebox{0.25cm}{!}{\protect\includegraphics{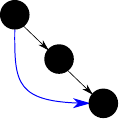}} vs. \protect\resizebox{0.25cm}{!}{\protect\includegraphics{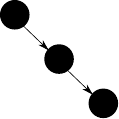}}. We observed that most people do not include these transitive connections. In particular, only two participants who worked on the open-ended task used explicit transitive connections to emphasize the relationship between events. However, even in the maps where they were used, they were scarce. Thus, in general, transitive connections were either not needed or participants had difficulty finding such connections in the first place. In contrast, the computer-generated maps from the original extraction algorithm \cite{keith2020maps} were able to easily extract explicit transitive connections and were well-evaluated by users. Therefore, we considered exploring whether including such explicit connections is useful. If so, using algorithms that can extract explicit transitive connections to emphasize specific relationships in narrative maps could be help analysts in their narrative sensemaking process. 

\item \textbf{Size}: The size of the map is inherently related to the length of the main story since longer main stories lead to bigger maps in general. Thus, the length of the main story acts as a proxy for map size. The number of events in the main story ranged from 6 to 25 events out of a total of 40 events in our data set. The mean number of events in the main story is 14.3 events and the median is 12.5 events.
\end{enumerate}

\begin{table*}[!htb]
\centering
\small
\resizebox{0.75\textwidth}{!}{%
\begin{tabular}{@{}cccccccccccc@{}}
\toprule
\textbf{Property}                                      & \textbf{Code}              & \textbf{D1} & \textbf{D2} & \textbf{D3} & \textbf{D4} & \textbf{D5} & \textbf{O1} & \textbf{O2} & \textbf{O3} & \textbf{O4} & \textbf{O5} \\ \midrule
\multirow{3}{*}{\textit{\textbf{Graph Structure}}}     & List \protect\resizebox{0.35cm}{!}{\protect\includegraphics{graphs/list.pdf}}                       &             &             &             &             & $\times$           &             &             &             &             & $\times$            \\
                                                       & Tree \protect\resizebox{0.35cm}{!}{\protect\includegraphics{graphs/tree.pdf}}                       &             &             & $\times$           & $\times$           &             & $\times$           & $\times$           &             &             &             \\
                                                       & DAG \protect\resizebox{0.25cm}{!}{\protect\includegraphics{graphs/dag.pdf}}                        & $\times$           & $\times$           &             &             &             &             &             & $\times$           & $\times$           &             \\ \midrule
\multirow{3}{*}{\textit{\textbf{Layout}}}              & Vertical (top-down) \protect\resizebox{0.25cm}{!}{\protect\includegraphics{graphs/vertical.pdf}}        &             & $\times$           & $\times$           & $\times$           &             & $\times$           & $\times$           & $\times$           & $\times$           &             \\
                                                       & Diagonal (left to right) \protect\resizebox{0.35cm}{!}{\protect\includegraphics{graphs/diagonal.pdf}}   & $\times$           &             &             &             &             &             &             &             &             &             \\
                                                       & Horizontal (left to right) \protect\resizebox{0.35cm}{!}{\protect\includegraphics{graphs/horizontal.pdf}} &             &             &             &             & $\times$           &             &             &             &             &  $\times$           \\ \midrule
\multirow{2}{*}{\textit{\textbf{Main Story Position}}} & Main Story First \protect\resizebox{0.30cm}{!}{\protect\includegraphics{graphs/main-first.pdf}}            &             & $\times$           & $\times$           & $\times$           & $\times$            & $\times$           &             &             & $\times$           & $\times$            \\
                                                       & Main Story Center \protect\resizebox{0.30cm}{!}{\protect\includegraphics{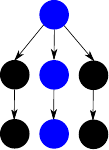}}          & $\times$           &             &             &             &             &             & $\times$           & $\times$           &             &             \\ \midrule

\multirow{2}{*}{\textit{\textbf{Source Nodes}}}      & Single \protect\resizebox{0.25cm}{!}{\protect\includegraphics{graphs/single-start.pdf}}               & $\times$            & $\times$           &             & $\times$           & $\times$           &             & $\times$           &             & $\times$           &             \\
                                                       & Multiple \protect\resizebox{0.25cm}{!}{\protect\includegraphics{graphs/multiple-start.pdf}}            &            &             & $\times$           &             &             & $\times$           &             & $\times$           &             & $\times$            \\ \midrule
\multirow{2}{*}{\textit{\textbf{Sink Nodes}}}        & Single \protect\resizebox{0.25cm}{!}{\protect\includegraphics{graphs/single-end.pdf}}                 &             & $\times$           &             &             & $\times$           &             &             &             &             &             \\
                                                       & Multiple \protect\resizebox{0.25cm}{!}{\protect\includegraphics{graphs/multiple-end.pdf}}              & $\times$           &             & $\times$           & $\times$           &             & $\times$           & $\times$           & $\times$           & $\times$           & $\times$            \\ \midrule
\multirow{2}{*}{\textit{\textbf{Connectivity}}}        & Connected \protect\resizebox{0.25cm}{!}{\protect\includegraphics{graphs/connected.pdf}}                 & $\times$             & $\times$            &             & $\times$             & $\times$           & $\times$             & $\times$             & $\times$             & $\times$             &             \\
                                                       & Disconnected \protect\resizebox{0.25cm}{!}{\protect\includegraphics{graphs/disconnected.pdf}}              &            &             & $\times$           &            &             &            &            &            &            & $\times$
                                                       \\ \midrule
\multirow{2}{*}{\textit{\textbf{Transitivity}}}        & Implicit \protect\resizebox{0.35cm}{!}{\protect\includegraphics{graphs/nontrans.pdf}}                 & $\times$             & $\times$            &   $\times$          & $\times$             & $\times$           & $\times$             & $\times$             &              &              &    $\times$         \\
                                                       & Explicit \protect\resizebox{0.35cm}{!}{\protect\includegraphics{graphs/transitivity.pdf}}              &            &             &            &            &             &            &            & $\times$           &  $\times$          & 
                                                       \\ \bottomrule
\end{tabular}%
}
\caption{Graph and Layout Properties for each participant in our study.}
\label{tab:layout}
\end{table*}

\subsection{Suggestions and Additional Features}
From the follow-up interviews, we also gathered a series of recommendations and suggestions for additional narrative map features. These suggestions were mostly oriented towards providing further support to the construction process and the subsequent use of the map. Setting aside the addition of basic functionalities, such as searching, highlighting, color-coding, or modifying the graph, as well as including more data, we summarize some of the key takeaways. First, participants mentioned the necessity of explanations in event connection (i.e., why are they connected?) and important events (i.e., why are they important?). Participants did not include any edge labels in their constructed maps, but they explained that they would prefer if maps created by other analysts included edge labels with explanations. Next, the participants mentioned the idea of getting automated recommendations on how to complete the map or expand it during the construction process, as this would make the construction process easier. Furthermore, maps should provide directions regarding the general topics or storylines in a specific part of the map (e.g., similar to section titles) and a way to focus on specific topics by zooming in with more details. Finally, events should be able to be merged if they are the same or above a certain similarity threshold, in order to reduce redundancy in the map.

\section{RQ2: Effects of Size and Transitivity}
\label{sec:rq4}
Based on our previous findings, we sought to explore the effects of size and transitivity on narrative maps. In particular, in RQ1 we found that the length of the main story in the analyst-generated maps had high variability, ranging from only 6 events to 25 events. Thus, we explored the effect of size on the utility of narrative maps. Moreover, in RQ1 we also found that most participants did not include explicit transitive connections. However, previous research has found that narrative maps that included these transitive connections were successful in terms of user evaluations \cite{keith2020maps}. Thus, we sought to compare maps with and without explicit transitive connections.

\subsection{Study Description}
To explore these characteristics, we performed a new experiment evaluating multiple combinations of sizes and the use of transitive connections. We opted to generate the maps computationally because this allows for easier scalability compared to manually generating maps for all the factor combinations in the experiment. Moreover, since our goal was to improve the pre-existing narrative maps design \cite{keith2020maps}, we generated a series of maps using pre-existing narrative extraction techniques. For the events, we used the same data set from RQ1.

\subsubsection{Narrative Extraction Algorithm}
We briefly describe the extraction algorithm that we used in this experiment. Our approach has two key parameters: the expected length of the main story ($K$), and the minimum coverage threshold.

We use an optimization method based on maximizing coherence---how much sense a storyline makes---subject to structural and topic coverage constraints with linear programming, following the approach by Keith and Mitra \cite{keith2020maps}. 

In particular, the structural constraints ensure that we obtain a directed acyclic graph with a single source and a single sink connected in chronological order through multiple storylines. The topic coverage constraints ensure that at least a certain percentage---based on the minimum coverage threshold---of the topics present in the data will be covered by the extracted narrative. 

Finally, our notion of coherence is based on similarity, under the logic that connected events should not drastically change their topics or contents throughout the narrative. Specifically, we compute the coherence value of joining two events by measuring their text similarity---based on an embedding representation---and their topical similarity---based on the same clusters from the coverage computation. 

\subsubsection{Map Size}
We extracted maps of different \textbf{sizes} based on the $K$ parameter of the extraction algorithm, which represents the expected length of the main story. We tested several levels: \textit{Small} ($K = 4$), \textit{Medium} ($K = 8$), \textit{Long} ($K = 12$), and \textit{Longer} ($K = 16$). A high value of $K$ leads to long and narrow maps, while a low value of $K$ results in shorter and wider maps. We generated all maps with a required minimum coverage of at least 50\% of the clusters found in the data (i.e., at least half of the relevant topics in the data should be covered by the map).

\subsubsection{Transitive Connections}
To study the effect of explicit transitive connections, we created maps with all their connections (normal maps) and maps with all explicit transitive connections removed (transitive reduced maps). To remove the extra connections from one of the base narrative maps we used \textit{transitive reduction}, an operation that removes edges on directed graphs while preserving its structure and important properties \cite{dubois2005transitive}. This operation is a way to reduce the complexity of large and dense graphs, which makes their layouts easier to read \cite{gansner2000open}. Thus, we would expect it to have a similar effect on narrative maps. We labeled maps using their \textit{Size} followed by a dash and \textit{N} for regular maps or \textit{T} for transitive reduced maps (e.g., Short-N).

\subsubsection{Evaluation Procedure}
\begin{figure}[!tbp]
    \centering
    \includegraphics[width=\columnwidth]{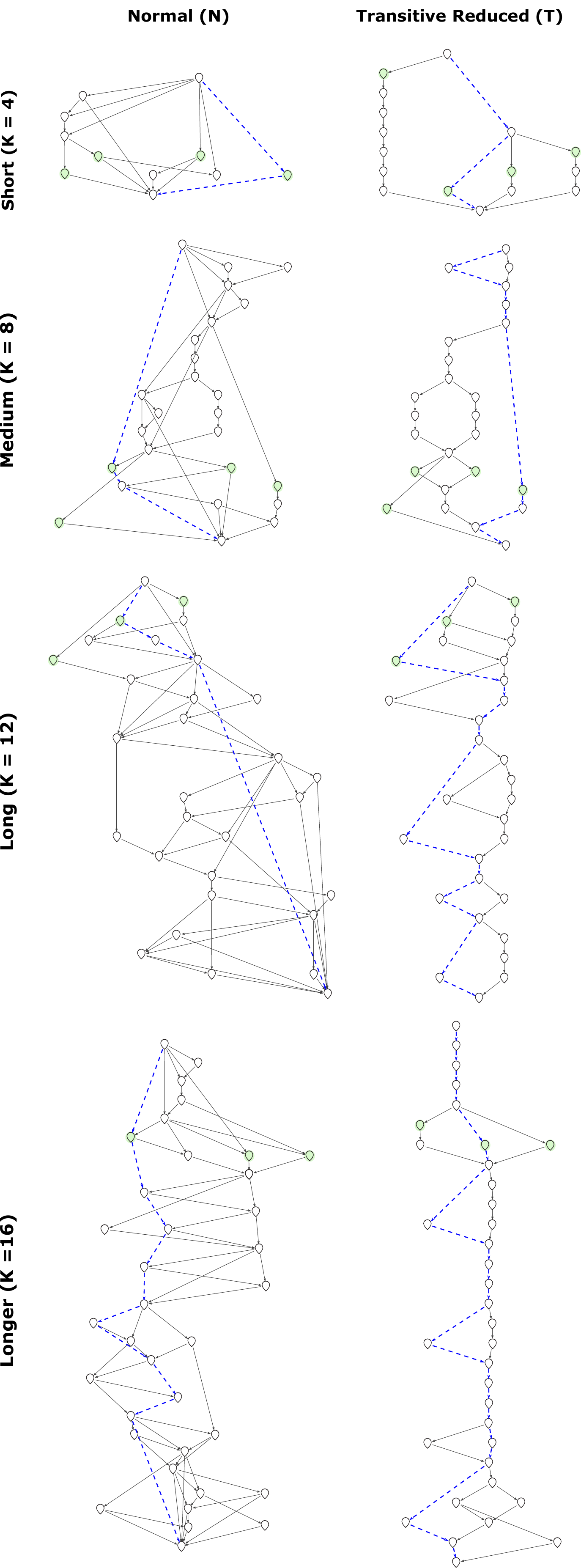}
    \caption{Overview of all the maps used in the evaluation procedure of RQ2. Normal maps (N) have more connections, allowing them to show more details at the cost of more complex layouts compared to their reduced counterparts (T).}
    \label{fig:minimaps}
\end{figure}

For evaluation purposes, we provided participants with a single map and asked them to complete narrative sensemaking tasks. We show a zoomed out overview of all the maps used in this evaluation in Figure \ref{fig:minimaps}.

In particular, we used the directed and open-ended tasks from our first experiment. The directed question could be answered by finding the main storyline in the extracted maps, while the second one could be answered by exploring the side storylines. Thus, these two tasks ensured that the participants had to make full use of the narrative map. We also included an evaluation questionnaire with ten 5-point Likert-scale questions. Then, we considered the percentage of favorable answers to evaluate the effectiveness of the narrative maps. We adapted the evaluation questionnaire used by Keith and Mitra \cite{keith2020maps}. This questionnaire considered multiple dimensions for the evaluation of narrative maps and adapted elements from similar procedures to evaluate the representation \cite{shahaf2010connecting,burkhard2005tube}, the metaphor \cite{burkhard2005tube,garcia2015visualisation}, and the visualization \cite{shamim2015evaluation}. We used a simplified version due to the stricter time constraints in this experiment. Nevertheless, this version covers all the main points of the original questionnaire (evaluating the underlying representation, the visualization itself, and the map metaphor). The relevant questions are listed below:

\begin{itemize}
\itemsep0em 
	    \item \textbf{Usefulness}: The map was helpful to answer the questions.	
	    \item \textbf{Coherence}: The map presents a coherent overview of the narrative.
	    \item \textbf{Relevance}: The map presents relevant information about the narrative.	
	    \item \textbf{Redundancy}: The map has redundant information. [Note the result for this question is reversed, as redundancy is a negative characteristic].
	    \item \textbf{User-friendliness}: The map is easy to understand.
	    \item \textbf{Comparability}: The map allows us to easily compare storylines.
	    \item \textbf{Completeness}: The amount of information on the map is appropriate to represent the narrative.
	    \item \textbf{Size}: The size of the map is appropriate to represent the narrative.
	    \item \textbf{Landmarks Metaphor}: The representative landmarks (green events) serve as an overview of all the stories in the narrative.
	    \item \textbf{Main Route Metaphor}: The main storyline (blue path) serves well as an overview of the most important events in the narrative.
\end{itemize}

\subsubsection{Study Participants}
Our design considered 91 potential subjects, which we randomly distributed among the factor combinations, ensuring that every combination had at least 11 subjects. The original sample consisted of 68 males and 29 females. The students were undergraduate students in a data analytics program. The participants had a lower level of experience compared to the participants of our first experiment, as their knowledge base consisted mostly of basic data analytics techniques. Nevertheless, most participants were able to complete the tasks. After filtering through blank and invalid responses, we had a total of 78 responses. Table \ref{fig:avg-results-exp3} shows the number of valid responses for each factor combination and the average effectiveness results.

\subsection{User Performance}
How well do users perform narrative sensemaking tasks with these narrative maps? To evaluate user performance, we identified a series of important \textit{high-level events} in the main story and the side stories. These high-level events are abstract representations of relevant events throughout the narrative. These high-level events were identified based on the narrative maps created for RQ1 as well as the follow-up interviews with participants. We evaluate user performance based on \textit{recall} (fraction of the high-level events that are successfully retrieved).

\begin{figure}[!htb]
\centering
    \includegraphics[width=0.80\columnwidth]{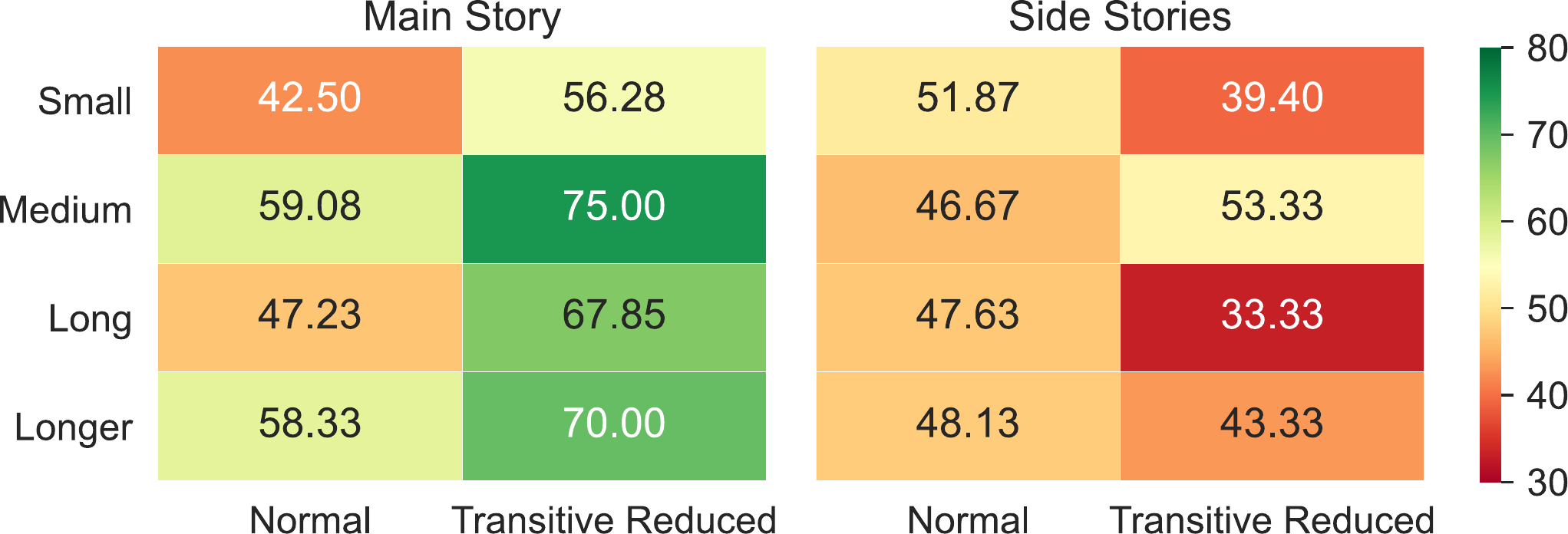}
\caption{Heat map showing the average recall for the main story and the side stories averaged over the issues.}
\label{fig:qual-analysis}
\end{figure}

In particular, the following high-level events that contributed to the US travel restrictions (i.e., the main story): the geographic spread of the virus, the reports on the virus's contagiousness, the death toll, and the worldwide responses. Moreover, we have the following notable high-level events for the side stories: the lockdown in China, the economic impacts, and the social impacts. We present the percentages of users that correctly identified these high-level events in the main story and the side stories are shown in Figure \ref{fig:qual-analysis}.

Medium-T (clean version shown in Figure \ref{fig:example}) had the highest recall of high-level events in both the main story and the side stories. Performing an ANOVA we find that the difference in main storylines is significant with respect to both map size and use of transitive connections ($p < 0.05$). In particular, Medium-T has the best performance in terms of recall. For the side storylines, the difference was not significant. The performance difference between the main story and the side stories shows that attempting to construct a single narrative map that covers both tasks is sub-optimal. Thus, using a series of task-specific maps rather than a single general map could lead to better results.

\subsection{User Evaluation Results}
\begin{figure}[!htb]
    \centering
    \includegraphics[width=0.48\textwidth]{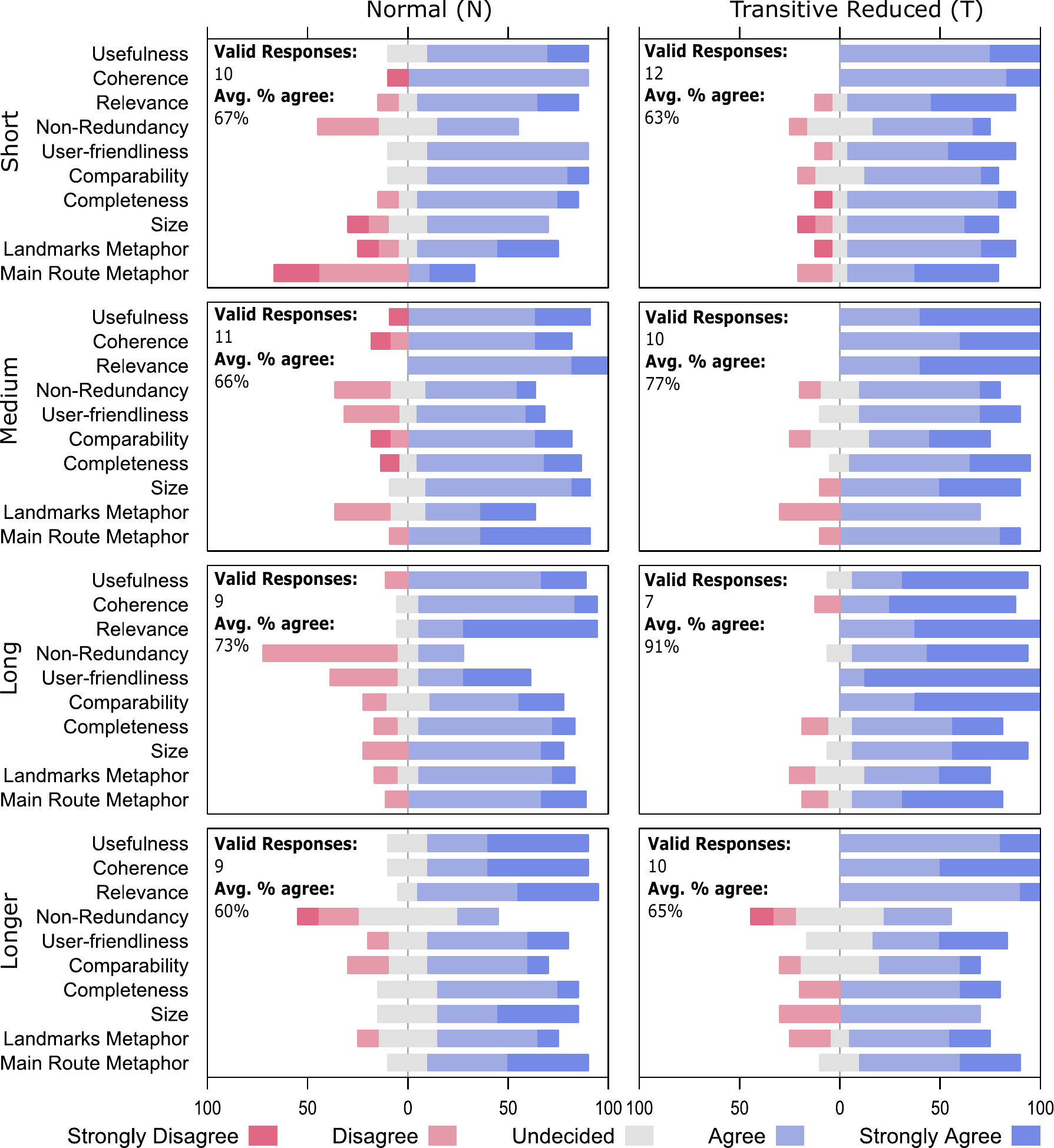}
    \caption{Percentage of favorable responses for each question and each \textit{size} and \textit{transitivity} combination. The best result was obtained by Long-T, followed by Medium-T, and then Long-N.}
    \label{fig:avg-results-exp3}
\end{figure}

How well do users evaluate these narrative maps in terms of effectiveness or utility? We show the evaluation questionnaire results in Figure \ref{fig:avg-results-exp3} and in Table \ref{tab:avg-results-exp3}. First, our best performing map is Long-T on most evaluation metrics, except for the metaphor-related metrics. On average, the second-best performing map is Medium-T and then Long-N. In particular, Long maps have the best performing results for all metrics.

\begin{table}[!htb]
\centering
\small
\resizebox{\columnwidth}{!}{%
\begin{tabular}{@{}ccccc@{}}
\toprule
                & \multicolumn{2}{c}{\textbf{Normal (N)}} & \multicolumn{2}{c}{\textbf{Transitive Reduced (T)}} \\ \midrule
\textbf{Size}   & Responses  & \% Agree & Responses        & \% Agree       \\ \midrule
\textbf{Short}  & 10               & 67.00            & 12                     & 63.33                  \\
\textbf{Medium} & 11               & 66.36            & 10                     & 77.00                  \\
\textbf{Long}   & 9                & 73.33            & 7                      & 91.43                  \\
\textbf{Longer} & 9                & 60.00            & 10                     & 65.00                  \\ \bottomrule
\end{tabular}%
}
\caption{Average percentage of favorable responses in our evaluation questionnaire for each \textit{size} and \textit{transitivity} combination. The best result was obtained by Long-T, followed by Medium-T, and then Long-N.}
\label{tab:avg-results-exp3}
\end{table}

The user preference for Long maps could be caused by their resemblance to timelines, which makes them more intuitive to use, while at the same time providing enough additional complexity to be useful as a narrative map. The tendency of users to prefer timeline-like structures could be related to the fact that timelines are the most basic and natural representation for narratives. Thus, users tend to prefer structures that are most familiar to them. Moreover, bigger maps are naturally able to contain more information than their smaller counterparts. However, Longer maps were not as well-received as Long maps, likely due to their unwieldy size and amount of content which made them impractical. 

Transitive reduced maps were better in all metrics except comparability (i.e., the ability to compare storylines), with an overall average of 72.8\% compared to 67.7\% for normal maps. For comparability, normal maps had 71.1\% favorable responses compared to only 56.4\% for transitive reduced maps. We hypothesize that this difference could be due to transitive reduction removing too many connections between storylines. Thus, by simplifying the map we lost important connections, making storyline comparison more difficult. Next, if we aggregate maps of the same size disregarding the effects of transitivity, Long maps have the advantage with an average of 81.3\% favorable responses, followed by Medium maps with 71.4\%.

Finally, we note that using transitive reduction on Short maps actually hurt the overall effectiveness (63\% compared to 67\%). This could mean that the extra connections present in the Small-N map were useful. Since smaller maps have fewer events, users benefited from knowing the connections between them. In contrast, bigger maps benefited from the use of transitive reduction to minimize complexity, at least up to a certain point, as Long-T performed better than Longer-T. Transitive reduction on Long maps had the highest positive effect on average user evaluation (from 73\% to 91\%). However, the benefits started to reduce as the map got bigger (from 60\% to only 65\% for Longer maps). Thus, this indicates that there is a sweet spot for the size of the map where transitive reduction has its greatest impact on effectiveness.

\section{Design Guidelines}
\label{sec:design}
\subsection{What makes a good narrative map?}
Based on our analysis of the results from all experiments, we present our narrative map design guidelines. These guidelines try to encapsulate the optimal design of narrative maps in the context of visual analytics and narrative sensemaking tasks. These recommendations seek to provide a general overview of what makes a ``good" narrative map. Table \ref{tab:summary} summarizes the design guidelines.

\rowcolors{2}{gray!25}{white}
\begin{table}[!htb]
\resizebox{\columnwidth}{!}{%
\begin{tabular}{@{}L{2cm}L{8.5cm}@{}}
\toprule
\rowcolor{white}
\textbf{Design Guidelines}              & \textbf{Summary}                                                                                                            \\ \midrule
Map Layout and Main Story Position      & According to our analysis, maps should be \textbf{vertical} \protect\resizebox{0.25cm}{!}{\protect\includegraphics{graphs/vertical.pdf}} with the main story displayed \textbf{first} \protect\resizebox{0.25cm}{!}{\protect\includegraphics{graphs/main-first.pdf}} (assuming a top-bottom and left-to-right order). In more general terms, we should use \textit{scrollytelling} and follow the reading order when designing narrative maps.                                                          \\
Starting and Ending Events              & From our results, we conclude that maps should have a \textbf{single starting event} \protect\resizebox{0.25cm}{!}{\protect\includegraphics{graphs/single-start.pdf}} (i.e., one source node) and potentially \textbf{multiple ending events} \protect\resizebox{0.25cm}{!}{\protect\includegraphics{graphs/multiple-end.pdf}} (i.e., many sink nodes) as their default design. However, we should allow for other combinations for flexibility, as depending on the task we might require different configurations of starting and ending events.                                                                            \\
Connectivity and Transitivity           & Narrative maps should be \textbf{connected} \protect\resizebox{0.25cm}{!}{\protect\includegraphics{graphs/connected.pdf}} (i.e., there should be no islands). Moreover, users prefer maps \textbf{without explicit transitive connections} \protect\resizebox{0.25cm}{!}{\protect\includegraphics{graphs/nontrans.pdf}}. The guiding principle in both of these guidelines is simplicity, as both of them help make the graph simpler for users.                                                                  \\
Map Size and Main Story Length          & Narrative maps have to balance the \textbf{trade-off between completeness and complexity}. In particular, bigger maps provide a more complete overview of the narrative, but they are more complex and harder to use. There is an optimal point in terms of size, but it depends on the specific data set. In general, map size should take into account the cognitive load that complex graphs place into analysts.                                                                                          \\
Cognitive Connections                   & Narrative maps should include \textbf{multiple types} of cognitive connections; ideally, they would leverage all types of connections. In general, we should replicate the rationale of analysts to connect events.                                                                                     \\
Event Selection & According to our analysis, maps should focus on \textbf{hard facts} and \textbf{impactful events}. In general, we should replicate the rationale of analysts to select events.\\

Graph Structure                         & \textbf{Structured approaches} (trees \protect\resizebox{0.45cm}{!}{\protect\includegraphics{graphs/tree.pdf}} or DAGs \protect\resizebox{0.25cm}{!}{\protect\includegraphics{graphs/dag.pdf}} ) are preferred over simple timelines. DAGs provide a more flexible and general approach over trees. We note that these structures align with the conceptual definition of narratives being complex systems of storylines.                                                  \\
Map Labeling                           & Narrative map edges should include \textbf{connection explanations} and important events should include a \textbf{justification} for their importance. In general, our narrative maps should include explanations to help users understand its elements.                                            \\
Edge Width and Length                           & Narrative map edges should include and express quantitative information as \textbf{edge widths}, but they should not rely on edge lengths, as distance is usually not a meaningful metric for graph layouts used with trees or DAGs.                                            \\
Storyline Presentation & Events should be partitioned into multiple \textbf{parallel storylines}, \textbf{labeled} and represented as \textbf{columns}. There should be a small number of important \textbf{inter-story connections}. In general, we should seek to maximize inter-story distance to ensure that the storylines are as distinct as possible. \\
Interactions                            & Relevant interactive features include adding a \textbf{recommendation system} for events in the narrative and the ability to \textbf{emphasize relevant keywords} based on user feedback. In general, it could be useful to include interactive AI techniques to improve narrative maps, such as using explainable AI and semantic interactions \cite{wenskovitch2020interactive}.                                           \\
Task-specific Maps & Narrative maps should be designed with a \textbf{specific target task} (e.g., directed and open-ended task). Users should have the option to create different maps based on their intended task. In general, we should specialize our narrative representations to the specific sensemaking task that we are trying to solve.\\ \bottomrule
\end{tabular}%
}
\caption{Summary of the design guidelines found in our analysis of results from RQ1 and RQ2.}
\label{tab:summary}
\end{table}

\textbf{Map Layout and Main Story Position:} Narrative maps should have a vertical layout. In visual storytelling terms, narrative maps should follow the \textit{scrollytelling} article design \cite{seyser2018scrollytelling}. Also, the main story should be shown first in the layout (assuming a top-bottom and left-to-right reading order). The influence of reading order is a known factor in how people perceive visual narratives \cite{segel2010narrative}. Despite the natural reading order of English being from left to right, participants predominantly used a top-down layout. The participants commented that vertical maps worked better since scrolling on a screen lends naturally to vertical layouts. Moreover, as our event headlines were horizontal, placing them vertically simulates natural reading in English. In particular, placing events in a horizontal fashion would have led to the narrative map being much wider and harder to use. Furthermore, we note that only participants that used timelines arranged the events in a horizontal fashion. This makes sense, as timelines are usually arranged in the same direction that people read (from left to right in our case) \cite{hendricks2015constructing}. However, more complex structures, such as trees and DAGs, do not necessarily follow this convention. Finally, we note that all our participants used standard computer screens. However, for smaller displays, such as cellphones or tablet screens, displaying maps horizontally might make more sense (i.e., closer to a \textit{slideshow} \cite{segel2010narrative} in visual storytelling terms). 

\textbf{Starting and Ending Events:} Narrative maps should have a single starting event and potentially multiple ending events by default. However, we should allow for other configurations, as depending on the task we might require a different combination of starting and ending events. We also note that our results might be biased because we gave participants a starting event in all tasks, but only the directed task had a predefined ending. However, most participants did not add their own starting events, even though many of them added their own ending events in addition to the predefined one when applicable.

\textbf{Connectivity and Transitivity:} Narrative maps should be connected and avoid using connections that can be induced by transitivity. During the creation process, almost no participants created disconnected components for their maps. This makes intuitive sense, since if something is so separate from the map that it must be its own disjointed component, then it is likely irrelevant to begin with. Regarding transitivity, both experiments showed that users generally prefer simpler maps where explicit transitive connections are omitted.

\textbf{Map Size and Main Story Length:} Based on our analyses, we found that the size of the map must balance the trade-off between completeness and complexity. Bigger maps are likely to present a more complete overview of the narrative at the expense of usability due to the increased complexity. Likewise, smaller maps are likely to be unable to cover the whole narrative, but are generally easier to use and understand due to their lower complexity. With our specific data set and evaluation context, we found that main stories should have length 12. In RQ1 the mean number of events in the main storyline was 14.3 events. However, these maps had high variability and there were many outliers. Thus, the median provides a better measure of the actual central point. In particular, the median was 12.5, which aligns with Long maps ($K = 12$) having the highest user evaluation in RQ2. Nevertheless, there are some caveats with this main story size. In particular, these results might depend on the size of the data set, the length of the timeline itself, and our resolution level. For example, if we are constructing a narrative that lasts a few years and we have daily updates, then 12 events might not be enough. Therefore, the results of this design guideline are specific to this data set. However, it might be applicable to similarly sized data sets, such as those used in other studies on sensemaking \cite{robinson2008collaborative,bradel2013analysts}. Finally, we note that based on the recall performance, an argument could be made towards favoring medium-sized maps. However, for the purposes of our design guidelines, we are interested in what users consider good in maps, rather than maximizing performance on this specific experiment.

\textbf{Event Selection:} Regarding the event selection criteria, narrative maps should be focused on hard facts and impactful events. Some events could be highlighted based on structural criteria, such as events that act as a hub node in the graph. Moreover, events should be selected from a variety of sources to create an unbiased narrative (e.g., including articles from left-leaning and right-leaning outlets to provide a politically balanced view of a narrative).

\textbf{Cognitive Connections:} Regarding connection types, rather than focusing on creating maps with a specific type of preferred connection, narrative map tools should allow the creation of different types of connections and provide an explanation of the type of connection (e.g., ``Common entity: China", ``Cause-effect relationship", or ``Same topic"). In general, based on the results and the use of different connection strategies by participants in our first experiment, a good narrative map would use a mix of different types of cognitive connections, rather than focusing on a specific one. 

In this respect, no existing tool in the literature handles all these cases. For example, the extraction algorithm for narrative maps uses similarity and topical connections \cite{keith2020maps}, but it does not include any cause-effect relationship or entity-based connections. In contrast, consider the Analyst's Workspace designed by Hossain et al. \cite{hossain2011helping}, which uses entity-based connections to generate storylines, but does not leverage topical information. As another example, consider the causal storytelling visualization technique developed by Choudhry et al. \cite{choudhry2020once}, which explicitly models causal relationships, but does not exploit other types of cognitive connections. Thus, we posit that there is a need to develop a narrative representation and extraction model that can leverage all these types of connections. 

Finally, we note the absence of \textit{citation-based connections} in the constructed narrative maps (i.e., A references B). This is a consequence of only considering headlines rather than the full articles, which could theoretically include links to previous articles in their body. However, even if we had the full text of the articles, we do not have HTML versions with hyperlinks available. Thus it would not be possible to find such type of connections with this data set. We note that this is a low-level type of connection, as it only requires analysts to detect the reference in the document, without necessarily analyzing it in more detail. However, it could turn into a higher level connection if the analysts detect why the reference was made in the first place. 

\textbf{Graph Structure:} Regarding the map structure, the map construction experiment showed that users preferred structured approaches (trees and DAGs) over simple timelines. However, there was no apparent preference between the tree-based approaches and the DAG-based approaches for the participants of RQ1. Thus, narrative maps should either use trees or DAGs for their underlying structure. However, we find that in theoretical terms, DAGs provide the most flexible representation. DAGs can be used to show divergent and convergent storylines, allowing for greater representation capabilities compared to a tree In particular, DAGs are also able to model timelines and trees. Thus, these representations could be interpreted as special cases of DAGs.

\textbf{Map Labeling:} The map should include labels that explain key components (i.e., connections and important events). We note that the analysts did not include explicit edge labels, but in the follow-up interviews, some participants were interested in having an explanation of why two events were joined together in the connections. This would aid other analysts in understanding the reasoning behind map connections. Thus, maps could include edge labels that provide qualitative information about the connection (e.g., ``Causal Connection") and they could also be color-coded for user convenience.  Similarly, participants also showed interest in explanations or justifications regarding why a specific event was deemed as important. Thus, maps should include additional labels providing information regarding connections and important events.

\textbf{Edge Width and Length}: In general, we note that the use of edge labels with numerical information can hurt usability, as discussed by Keith and Mitra, a better alternative is to instead show such information via edge widths \cite{keith2020maps}. However, in our experiments, we found that most participants did not include explicit weight information for the connections nor any other type of quantitative labeling. Moreover, they did not use arrows of different width or length to denote connection strength. Nevertheless, edge widths could be used to provide quantitative information, such as connection strength (e.g., coherence). In contrast, edge length would be harder to use for this purpose, as the graph layouts used in the construction of narrative maps are not directly based on the concept of distance, but on hierarchies and levels generated by trees and DAGs.

\textbf{Storyline Presentation}: Events should be partitioned into parallel storylines. The storylines should be labeled and represented as columns. Moreover, there should be a small number of important inter-story connections joining them. We note that all participants separated their events into clearly defined parallel storylines of varying lengths and labeled their storylines. The labels were based on the general topic or issue presented in the storyline (e.g., prevention or economic impacts). In most cases, these storylines had few inter-story connections between them. Participants also labeled their storylines. 

\textbf{Interactions:} In addition to all the structural and layout information obtained from our first experiment, we also obtained some insight towards potential useful interactions for a narrative maps tool. Beyond basic surface-level interactions, such as zooming, moving elements, changing layouts, and adding/deleting elements, participants were interested in obtaining recommendations on which events to add next as they constructed the map. In general, users expressed interest in a recommendation system to enrich the narrative map. Moreover, this system could be extended to also detect missing events in existing storylines. We note that there are other interactions or approaches that could be beneficial when confronted with a larger corpus of related articles. For example, using similarity of the articles to find related story pieces or citations and cross-references to detect other relevant articles. Such dependencies could be offered so that analysts could drag and drop new events into existing paths.

A close analogue in the literature to the narrative maps method is the metro maps approach developed by Shahaf et al. \cite{shahaf2012trains}. This visualization tool incorporates its own narrative representation and extraction algorithm. In particular, it incorporates user feedback through the selection of important tags (i.e., selecting relevant words according to the user's interests). A similar approach could be used to incorporate user feedback into narrative maps. Lastly, our participants mentioned the idea of incorporating user feedback through keywords as a way to obtain a more relevant map. This could be implemented by emphasizing events based on input keywords on a search bar or highlighted words by the user. In general, it could be useful to include interactive AI techniques to improve narrative maps, such as using explainable AI and semantic interactions

\textbf{Task-specific Maps:} Narrative maps should be designed with a specific target task. We found that using a single narrative map to attempt answering both the directed and open-ended tasks was sub-optimal. The current extraction algorithm focuses on the directed task, thus leading to lower performance in recognizing important side stories. While a combined map for both tasks can provide an appropriate overview, it would be better to create task-specific maps. For example, for the open-ended task, we could generate a map without a fixed ending event and with high topic coverage, as this would likely lead to a narrative map that explored multiple outcomes.

\section{Discussion}
\subsection{Visual Storytelling and Narrative Maps}
Regarding the use of visual storytelling techniques with narrative maps, we discuss some concepts, taken from the work of Segel and Heer \cite{segel2010narrative}, that could be potentially useful in the design of an interactive visualization tool for narrative maps. 

Regarding \textit{visual narrative} elements. We note that narrative maps should guide viewers to explore paths in the visualization through the use of \textit{visual highlighting} (e.g., color, size, boldness). In practice, this would require highlighting the main storyline, but there should also be clear indications for side stories. The ability to perform \textit{close-ups} or \textit{zooming} into relevant map sections is also important.

Regarding \textit{messaging}, narrative maps already include the \textit{headlines} of the events as core elements of the narrative. Nevertheless, there are other messaging tools from storytelling that could be used. For example, \textit{annotations}, such as edge labels, storyline names, or other macro-structures names (e.g., clusters) to the narrative map could prove useful as well. The inclusion of a \textit{summary} could also be a useful feature, as it would be able to provide additional context and a brief overview of the content of the map. 

Regarding \textit{interactivity} elements, narrative maps should also consider including a \textit{details-on-demand} feature, either by mousing-over an event on the graph or by clicking on them. Such a feature could open a special details tab, containing information such as the full article, a snapshot of the original publication, or even a list of related articles. It could also be useful to include a \textit{timeline slider} element, as it could allow users to change the scope of the visualized narrative to a different time window.  Moreover, it should be possible to perform \textit{filtering}, \textit{selection}, and \textit{searching} over the events of the narrative.

\subsection{Influence of Analyst Background and Experience}
First, we note that the analysts were not working professional analysts, but were student analysts-in-training. Thus, their specific sensemaking strategies might be influenced by their lower level of experience compared to real analysts. Moreover, if analysts were familiar with structured analytic techniques \cite{pherson2020structured}, such as the generic narrative space model \cite{sappelsa2013generic} or other methods, they might affect their sensemaking process, as these techniques provide ways to develop compelling narrative rationales \cite{pherson2020structured,cheng2019explaining}. Nevertheless, previous work has shown that studies with real analysts and with students have similar findings and implications \cite{kang2010can}. Exploring the influence of specific analyst experience and is left as future work. More specifically, future work should include the study of more cases with professional analysts. 

\subsection{Influence of the Data Set and Task Choice}
Regarding the data set, we note that the use of a current topic such as COVID-19 might have influenced the results, as participants could have been influenced heavily by their own experiences with the pandemic. Moreover, the data set was relatively small, a limitation imposed due to time constraints. The data set size could make it difficult to scale the detected strategies or results to larger data sets, which, for example, could place more emphasis on the foraging steps of the sensemaking loop or require more complex narrative map structures. Regardless of these issues, the COVID-19 data set should still provide valuable insight into the synthesis loop part of the sensemaking process. Moreover, the data set size is in line with related works \cite{robinson2008collaborative, bradel2013analysts} that use intelligence analysis data sets \cite{hughes2003discovery}, such as \textit{The Sign of the Crescent} data set (41 documents) or the \textit{Atlantic Storm} data set (47 documents) to understand the analyst sensemaking process.

It should also be noted that experience and prior knowledge might heavily influence the work done by participants, especially due to the use of a recent and high-profile topic such as COVID-19. In particular, participants had different levels of expertise on the topic and were able to bring insights from their own knowledge and experiences. Specifically, in the RQ1 experiment, we note that only four analysts made explicit remarks on how they used domain knowledge in their construction process. However, the other six analysts might have drawn on this knowledge implicitly without properly acknowledging it.

In addition to this, we note that the specific choice of starting and ending events in the directed task also influences the construction of the map and what is considered part of the main storyline or a side story. For example, when trying to find the connection between the initial outbreak and travel restrictions, it is unlikely that documents relating to oil prices are directly part of the main story. However, if the question required connecting the dots between the initial outbreak and the economic impacts it would make more sense as part of the main storyline.

Finally, we note that both of the tasks used in this study represent simplified and constrained versions of what analysts would do in a real-world setup, but they still provide valuable insights into the general narrative sensemaking process. Nevertheless, as these tasks do not capture the full sensemaking process, caution should be exercised when attempting to generalize these conclusions, especially as higher complexity tasks might yield different kinds of strategies or structures. 

\subsection{Sensemaking Process}
We note that much of the evaluation process rediscovers parts of the larger sensemaking process. However, in this article, we focus exclusively on how the synthesis loop of the sensemaking process applies to narrative maps. Thus, the results are only applicable to this scope. Future work could address how other types of sensemaking strategies or tools compare against narrative maps. 

Furthermore, it would be useful for future work to do multiple evaluations with data sets with different characteristics and analysts with different levels of experience. Such work could ask analysts to create a narrative map based on their own analytical work that they have previously completed as part of their regular practice, as opposed to using a specific toy data set, although such an approach would have several more variables to account for, requiring careful experimental design. However, there would be value in drawing lessons and guidelines from a more diverse set of analytic problems, as this would also provide information on where and how narrative maps could be best applied.

\subsection{Interaction with Other Guidelines}
We note that our proposed guidelines focus on the design of the narrative maps, but any implementation of an interactive tool for narrative maps should consider general visualization principles and design guidelines, such as the visualization mantra \cite{shneiderman2003eyes}: ``overview first, filtering and selection, then details on demand''. For example, by presenting users with an overview of the map at first, then letting them zoom to specific storylines or components of the map, and then providing specific details about the events as needed.

\subsection{Limitations}
Our work is not without limitations. First, there is an unbalanced number of participants in the two experiments, the first one has 10 subjects, based on the methodology of Bradel et al. \cite{bradel2013analysts}, while the second had 78 valid responses. Due to the qualitative nature of the first experiment and the need to understand the construction strategies in depth, it was necessary to use a much smaller sample size. In contrast, the second experiment did not require such a level of detail, making it much simpler to scale up. However, we note that the difference in sample size makes comparing results between these experiments more complex.

Regarding the limitations of the RQ1 experiment, we note that we conducted interviews with only a handful of analysts (10). While the number was small, all participants had a background in intelligence analysis. They also spanned a variety of majors and had reasonable gender representation (6 females and 4 males). Nevertheless, even with 10 participants, we were able to observe diverse strategies and structures for narrative map construction. 

Regarding the limitations of the RQ2 experiment, we first note that each factor combination had a different response rate, as not all participants completed the assigned tasks. Nevertheless, the general trend still provided useful insight towards how to design narrative maps. Another issue was the lack of experience of the participants; however, the data set was small enough and the questions were designed to be simple so even non-expert users could answer them. Finally, we note that this experiment lacks an explicit baseline, such as a basic timeline or similar representation.

\section{Conclusions}
\label{sec:conclusions}
We studied how analysts construct narrative maps and the characteristics of these maps. In particular, our user study detected 7 types of cognitive connections. In particular, we have shown the importance of topical and causal relationships in the construction of narrative maps, as these were the most common high-level connections in the user-generated maps. 

In terms of strategies, we found three major ways to construct maps. Each one of these strategies can be the basis of a new extraction algorithm. Furthermore, in terms of the structure of the map, we saw an even distribution between tree-like maps and DAG-like maps. Regarding layout, we found that most users preferred a vertical top-down layout (i.e., \textit{scrollytelling}), with the main story shown first. We also evaluated the effect of map size and transitivity, finding that users preferred long maps without transitive connections. 

All these results led to a series of design guidelines for narrative maps. These guidelines can be used in the design of new extraction algorithms and interactive visualization tools. Future work will deal with the implementation of such algorithms and tools, as well as their evaluation based on the insights gathered in this work.

Future work could explore how strategies differ when applied to different domains, data set sizes, and analyst experience. In particular, it would be useful to consider how previous analyst training (e.g., experience with structured analytic techniques) could influence the construction strategies or the narrative map structures.

Finally, as mentioned before, the overarching goal of our study was to improve the design of narrative maps \cite{keith2020maps}. Thus, by extracting these design guidelines and understanding the narrative sensemaking process, we have provided the basis for future improvements of the narrative map model. Thus, future work should focus on using these findings to improve narrative maps and the associated extraction algorithms.


\begin{thebibliography}{10}
\providecommand{\url}[1]{\texttt{#1}}
\providecommand{\urlprefix}{URL }
\expandafter\ifx\csname urlstyle\endcsname\relax
  \providecommand{\doi}[1]{DOI:\discretionary{}{}{}#1}\else
  \providecommand{\doi}{DOI:\discretionary{}{}{}\begingroup
  \urlstyle{rm}\Url}\fi
\providecommand{\eprint}[2][]{\url{#2}}

\bibitem{halverson2011master}
Halverson J, Corman S and Goodall HL.
\newblock \emph{Master narratives of Islamist extremism}.
\newblock 175 5th Ave., New York, NY, USA: Springer, 2011.

\bibitem{abbott2008cambridge}
Abbott HP.
\newblock \emph{The Cambridge introduction to narrative}.
\newblock One Liberty Plaza, New York, NY, USA: Cambridge University Press,
  2008.

\bibitem{finlayson2013military}
Finlayson MA and Corman SR.
\newblock The military interest in narrative.
\newblock \emph{Sprache und Datenverarbeitung} 2013; 37(1-2): 173--191.

\bibitem{hossain2011helping}
Hossain MS, Andrews C, Ramakrishnan N et~al.
\newblock Helping intelligence analysts make connections.
\newblock In \emph{Workshops at the Twenty-Fifth AAAI Conference on Artificial
  Intelligence (2011)}. pp. 1--10.

\bibitem{hossain2012connecting}
Hossain MS, Gresock J, Edmonds Y et~al.
\newblock Connecting the dots between pubmed abstracts.
\newblock \emph{PloS one} 2012; 7(1): e29509.

\bibitem{choudhry2020once}
Choudhry A, Sharma M, Chundury P et~al.
\newblock Once upon a time in visualization: Understanding the use of textual
  narratives for causality.
\newblock \emph{IEEE Transactions on Visualization and Computer Graphics} 2021;
  27(2): 1332--1342.
\newblock \doi{10.1109/TVCG.2020.3030358}.

\bibitem{segel2010narrative}
Segel E and Heer J.
\newblock Narrative visualization: Telling stories with data.
\newblock \emph{IEEE Transactions on Visualization and Computer Graphics} 2010;
  16(6): 1139--1148.
\newblock \doi{10.1109/TVCG.2010.179}.

\bibitem{tong2018storytelling}
Tong C, Roberts R, Borgo R et~al.
\newblock Storytelling and visualization: An extended survey.
\newblock \emph{Information} 2018; 9(3).
\newblock \doi{10.3390/info9030065}.

\bibitem{riche2018data}
Riche NH, Hurter C, Diakopoulos N et~al.
\newblock \emph{Data-driven storytelling}.
\newblock CRC Press, 2018.

\bibitem{keith2020maps}
Keith~Norambuena BF and Mitra T.
\newblock Narrative maps: An algorithmic approach to represent and extract
  information narratives.
\newblock \emph{Proc ACM Hum-Comput Interact} 2021; 4(CSCW3).
\newblock \doi{10.1145/3432927}.

\bibitem{bradel2015big}
Bradel L, Wycoff N, House L et~al.
\newblock Big text visual analytics in sensemaking.
\newblock In \emph{2015 Big Data Visual Analytics (BDVA)}. pp. 1--8.
\newblock \doi{10.1109/BDVA.2015.7314287}.

\bibitem{endert2014human}
Endert A, Hossain MS, Ramakrishnan N et~al.
\newblock The human is the loop: new directions for visual analytics.
\newblock \emph{Journal of Intelligent Information Systems} 2014; 43(3):
  411--435.

\bibitem{cook2005illuminating}
Cook KA and Thomas JJ.
\newblock Illuminating the path: The research and development agenda for visual
  analytics.
\newblock Technical report, Pacific Northwest National Laboratory (PNNL),
  Richland, WA, US, 2005.

\bibitem{pirolli2005sensemaking}
Pirolli P and Card S.
\newblock The sensemaking process and leverage points for analyst technology as
  identified through cognitive task analysis.
\newblock In \emph{Proceedings of international conference on intelligence
  analysis (2005)}, volume~5. McLean, VA, USA: International Conference on
  Intelligence Analysis, pp. 2--4.

\bibitem{kang2009evaluating}
{Kang} Y, {Gorg} C and {Stasko} J.
\newblock Evaluating visual analytics systems for investigative analysis:
  Deriving design principles from a case study.
\newblock In \emph{2009 IEEE Symposium on Visual Analytics Science and
  Technology}. New York, NY, USA: IEEE, pp. 139--146.
\newblock \doi{10.1109/VAST.2009.5333878}.

\bibitem{wright2006sandbox}
Wright W, Schroh D, Proulx P et~al.
\newblock The sandbox for analysis: Concepts and methods.
\newblock In \emph{Proceedings of the SIGCHI Conference on Human Factors in
  Computing Systems}. CHI 20606, New York, NY, USA: Association for Computing
  Machinery.
\newblock ISBN 1595933727, p. 801–810.
\newblock \doi{10.1145/1124772.1124890}.

\bibitem{ho2001towards}
Ho J and Tang R.
\newblock Towards an optimal resolution to information overload: An infomediary
  approach.
\newblock In \emph{Proceedings of the 2001 International ACM SIGGROUP
  Conference on Supporting Group Work}. GROUP 2001, New York, NY, USA:
  Association for Computing Machinery.
\newblock ISBN 1581132948, p. 91–96.
\newblock \doi{10.1145/500286.500302}.

\bibitem{shahaf2010connecting}
Shahaf D and Guestrin C.
\newblock Connecting the dots between news articles.
\newblock In \emph{Proceedings of the 16th ACM SIGKDD International Conference
  on Knowledge Discovery and Data Mining}. KDD 2010, New York, NY, USA:
  Association for Computing Machinery.
\newblock ISBN 9781450300551, p. 623–632.
\newblock \doi{10.1145/1835804.1835884}.

\bibitem{bex2006anchored}
Bex F, Prakken H and Verhey B.
\newblock Anchored narratives in reasoning about evidence.
\newblock In \emph{Proceedings of the 2006 Conference on Legal Knowledge and
  Information Systems: JURIX 2006: The Nineteenth Annual Conference}, volume
  152. IOS Press.
\newblock ISBN 158603698X, p. 11–20.

\bibitem{sappelsa2013generic}
Sappelsa L, Parunak HVD and Brueckner S.
\newblock The generic narrative space model as an intelligence analysis tool.
\newblock \emph{American Intelligence Journal} 2013; 31(2): 69--78.

\bibitem{norambuena2021narrative}
Keith~Norambuena BF, Mitra T and North C.
\newblock Narrative sensemaking: Strategies for narrative maps construction,
  2021.

\bibitem{puckett2016narrative}
Puckett K.
\newblock \emph{Narrative theory}.
\newblock One Liberty Plaza, New York, NY, USA: Cambridge University Press,
  2016.

\bibitem{hullman2013deeper}
Hullman J, Drucker S, Henry~Riche N et~al.
\newblock A deeper understanding of sequence in narrative visualization.
\newblock \emph{IEEE Transactions on Visualization and Computer Graphics} 2013;
  19(12): 2406--2415.
\newblock \doi{10.1109/TVCG.2013.119}.

\bibitem{hullman2017finding}
Hullman J, Kosara R and Lam H.
\newblock Finding a clear path: Structuring strategies for visualization
  sequences.
\newblock \emph{Computer Graphics Forum} 2017; 36(3): 365--375.
\newblock \doi{https://doi.org/10.1111/cgf.13194}.

\bibitem{hullman2013contextifier}
Hullman J, Diakopoulos N and Adar E.
\newblock Contextifier: Automatic generation of annotated stock visualizations.
\newblock In \emph{Proceedings of the SIGCHI Conference on Human Factors in
  Computing Systems}. CHI 2013, New York, NY, USA: Association for Computing
  Machinery.
\newblock ISBN 9781450318990, p. 2707–2716.
\newblock \doi{10.1145/2470654.2481374}.

\bibitem{gao2014newsviews}
Gao T, Hullman JR, Adar E et~al.
\newblock Newsviews: An automated pipeline for creating custom
  geovisualizations for news.
\newblock In \emph{Proceedings of the SIGCHI Conference on Human Factors in
  Computing Systems}. CHI 2014, New York, NY, USA: Association for Computing
  Machinery.
\newblock ISBN 9781450324731, p. 3005–3014.
\newblock \doi{10.1145/2556288.2557228}.

\bibitem{akaishi2007narrative}
Akaishi M, Yoshikiyo K, Satoh K et~al.
\newblock Narrative based topic visualization for chronological data.
\newblock In \emph{2007 11th International Conference Information Visualization
  (IV 2007)}. pp. 139--144.
\newblock \doi{10.1109/IV.2007.80}.

\bibitem{wang2015socially}
Wang L, Cardie C and Marchetti G.
\newblock Socially-informed timeline generation for complex events.
\newblock In \emph{Proceedings of the 2015 Conference of the North {A}merican
  Chapter of the ACL: Human Language Technologies}. Denver, Colorado: ACL, pp.
  1055--1065.

\bibitem{shahaf2012trains}
Shahaf D, Guestrin C and Horvitz E.
\newblock Trains of thought: Generating information maps.
\newblock In \emph{Proceedings of the 21st International Conference on World
  Wide Web}. WWW 2012, New York, NY, USA: Association for Computing Machinery.
\newblock ISBN 9781450312295, p. 899–908.
\newblock \doi{10.1145/2187836.2187957}.

\bibitem{liu2017growing}
Liu B, Niu D, Lai K et~al.
\newblock Growing story forest online from massive breaking news.
\newblock In \emph{Proceedings of the 2017 ACM on Conference on Information and
  Knowledge Management}. CIKM 2017, New York, NY, USA: Association for
  Computing Machinery.
\newblock ISBN 9781450349185, p. 777–785.
\newblock \doi{10.1145/3132847.3132852}.

\bibitem{nallapati2004event}
Nallapati R, Feng A, Peng F et~al.
\newblock Event threading within news topics.
\newblock In \emph{Proceedings of the Thirteenth ACM International Conference
  on Information and Knowledge Management}. CIKM ’04, New York, NY, USA: ACM.
\newblock ISBN 1581138741, p. 446–453.

\bibitem{kim2011topic}
Kim D and Oh A.
\newblock Topic chains for understanding a news corpus.
\newblock In Gelbukh A (ed.) \emph{Computational Linguistics and Intelligent
  Text Processing (2011)}. Berlin, Heidelberg: Springer Berlin Heidelberg.
\newblock ISBN 978-3-642-19437-5, pp. 163--176.

\bibitem{zhou2017survey}
Zhou H, Yu H, Hu R et~al.
\newblock A survey on trends of cross-media topic evolution map.
\newblock \emph{Knowledge-Based Systems} 2017; 124: 164--175.
\newblock \doi{https://doi.org/10.1016/j.knosys.2017.03.009}.

\bibitem{faloutsos2004fast}
Faloutsos C, McCurley KS and Tomkins A.
\newblock Fast discovery of connection subgraphs.
\newblock In \emph{Proceedings of the Tenth ACM SIGKDD International Conference
  on Knowledge Discovery and Data Mining}. KDD 2004, New York, NY, USA:
  Association for Computing Machinery.
\newblock ISBN 1581138881, p. 118–127.
\newblock \doi{10.1145/1014052.1014068}.

\bibitem{soni2014modeling}
Soni S, Mitra T, Gilbert E et~al.
\newblock Modeling factuality judgments in social media text.
\newblock In \emph{Proceedings of the 52nd Annual Meeting of the ACL (Volume 2:
  Short Papers)}. Baltimore, Maryland: ACL, pp. 415--420.

\bibitem{shahaf2013information}
Shahaf D, Yang J, Suen C et~al.
\newblock Information cartography: Creating zoomable, large-scale maps of
  information.
\newblock In \emph{Proceedings of the 19th ACM SIGKDD International Conference
  on Knowledge Discovery and Data Mining}. KDD 2013, New York, NY, USA:
  Association for Computing Machinery.
\newblock ISBN 9781450321747, p. 1097–1105.
\newblock \doi{10.1145/2487575.2487690}.

\bibitem{baikadi2011towards}
Baikadi A, Goth J, Mitchell C et~al.
\newblock Towards a computational model of narrative visualization.
\newblock \emph{Proceedings of the AAAI Conference on Artificial Intelligence
  and Interactive Digital Entertainment} 2011; 7(2): 2--9.

\bibitem{kelter2004representing}
Kelter S, Kaup B and Claus B.
\newblock Representing a described sequence of events: a dynamic view of
  narrative comprehension.
\newblock \emph{Journal of Experimental Psychology: Learning, Memory, and
  Cognition} 2004; 30(2): 451.

\bibitem{yan2011evolutionary}
Yan R, Wan X, Otterbacher J et~al.
\newblock Evolutionary timeline summarization: A balanced optimization
  framework via iterative substitution.
\newblock In \emph{Proceedings of the 34th International ACM SIGIR Conference
  on Research and Development in Information Retrieval}. SIGIR 2011, New York,
  NY, USA: Association for Computing Machinery.
\newblock ISBN 9781450307574, p. 745–754.
\newblock \doi{10.1145/2009916.2010016}.

\bibitem{shahaf2012connecting}
Shahaf D and Guestrin C.
\newblock Connecting two (or less) dots: Discovering structure in news
  articles.
\newblock \emph{ACM Trans Knowl Discov Data} 2012; 5(4).
\newblock \doi{10.1145/2086737.2086744}.

\bibitem{ansah2019graph}
Ansah J, Liu L, Kang W et~al.
\newblock A graph is worth a thousand words: Telling event stories using
  timeline summarization graphs.
\newblock In \emph{The World Wide Web Conference}. WWW ’19, New York, NY,
  USA: ACM.
\newblock ISBN 9781450366748, p. 2565–2571.

\bibitem{yang2009discovering}
Yang CC, Shi X and Wei CP.
\newblock Discovering event evolution graphs from news corpora.
\newblock \emph{IEEE Transactions on Systems, Man, and Cybernetics - Part A:
  Systems and Humans} 2009; 39(4): 850--863.
\newblock \doi{10.1109/TSMCA.2009.2015885}.

\bibitem{zhou2017emmbtt}
Zhou P, Wu B and Cao Z.
\newblock Emmbtt: A novel event evolution model based on tfxief and tdc in
  tracking news streams.
\newblock In \emph{2017 IEEE Second International Conference on Data Science in
  Cyberspace (DSC)}. pp. 102--107.
\newblock \doi{10.1109/DSC.2017.53}.

\bibitem{xu2013summarizing}
Xu S, Wang S and Zhang Y.
\newblock Summarizing complex events: a cross-modal solution of storylines
  extraction and reconstruction.
\newblock In \emph{Proceedings of the 2013 Conference on Empirical Methods in
  Natural Language Processing}. Seattle, Washington, USA: ACL, pp. 1281--1291.

\bibitem{shahaf2013metro}
Shahaf D, Guestrin C and Horvitz E.
\newblock Metro maps of information.
\newblock \emph{SIGWEB Newsletter} 2013; Spring.

\bibitem{bradel2013analysts}
{Bradel} L, {Self} JZ, {Endert} A et~al.
\newblock How analysts cognitively “connect the dots”.
\newblock In \emph{2013 IEEE International Conference on Intelligence and
  Security Informatics}. New York, NY, USA: IEEE, pp. 24--26.
\newblock \doi{10.1109/ISI.2013.6578780}.

\bibitem{robinson2008collaborative}
Robinson AC.
\newblock Collaborative synthesis of visual analytic results.
\newblock In \emph{2008 IEEE Symposium on Visual Analytics Science and
  Technology}. pp. 67--74.
\newblock \doi{10.1109/VAST.2008.4677358}.

\bibitem{andrews2010space}
Andrews C, Endert A and North C.
\newblock Space to think: Large high-resolution displays for sensemaking.
\newblock In \emph{Proceedings of the SIGCHI Conference on Human Factors in
  Computing Systems}. CHI 2010, New York, NY, USA: Association for Computing
  Machinery.
\newblock ISBN 9781605589299, p. 55–64.
\newblock \doi{10.1145/1753326.1753336}.

\bibitem{andrews2012analyst}
Andrews C and North C.
\newblock Analyst's workspace: An embodied sensemaking environment for large,
  high-resolution displays.
\newblock In \emph{2012 IEEE Conference on Visual Analytics Science and
  Technology (VAST)}. pp. 123--131.
\newblock \doi{10.1109/VAST.2012.6400559}.

\bibitem{wenskovitch2020examination}
Wenskovitch J and North C.
\newblock An examination of grouping and spatial organization tasks for
  high-dimensional data exploration.
\newblock \emph{IEEE Transactions on Visualization and Computer Graphics} 2021;
  27(2): 1742--1752.
\newblock \doi{10.1109/TVCG.2020.3028890}.

\bibitem{haider2017analysts}
Haider JD, Seidler P, Pohl M et~al.
\newblock How analysts think: Sense-making strategies in the analysis of
  temporal evolution and criminal network structures and activities.
\newblock \emph{Proceedings of the Human Factors and Ergonomics Society Annual
  Meeting} 2017; 61(1): 193--197.
\newblock \doi{10.1177/1541931213601532}.

\bibitem{endert2012clustering}
Endert A, Fox S, Maiti D et~al.
\newblock The semantics of clustering: Analysis of user-generated
  spatializations of text documents.
\newblock In \emph{Proceedings of the International Working Conference on
  Advanced Visual Interfaces}. AVI 2012, New York, NY, USA: Association for
  Computing Machinery.
\newblock ISBN 9781450312875, p. 555–562.
\newblock \doi{10.1145/2254556.2254660}.

\bibitem{camargo2013manual}
Camargo RT, Agostini V, Di~Felippo A et~al.
\newblock Manual typification of source texts and multi-document summaries
  alignments.
\newblock \emph{Procedia-Social and Behavioral Sciences} 2013; 95: 498--506.

\bibitem{khandkar2009open}
Khandkar SH.
\newblock Open coding.
\newblock \emph{University of Calgary} 2009; 23: 2009.

\bibitem{norambuenaevaluating}
Norambuena BK, Horning M and Mitra T.
\newblock Evaluating the inverted pyramid structure through automatic 5w1h
  extraction and summarization.
\newblock In \emph{Proc. of the 2020 Computation + Journalism Symposium}.
  Boston, MA, USA: Computation + Journalism 2020, pp. 1--7.

\bibitem{baikadi2012towards}
Baikadi A and Cardona-Rivera RE.
\newblock Towards finding the fundamental unit of narrative: A proposal for the
  narreme.
\newblock In \emph{The Third Workshop on Computational Models of Narrative}.
  MIT CSAIL, Boston, MA, USA: MIT, pp. 44--46.

\bibitem{de2005news}
De~Vreese CH.
\newblock News framing: Theory and typology.
\newblock \emph{Information design journal \& document design} 2005; 13(1):
  51--62.

\bibitem{dubois2005transitive}
Dubois V and Bothorel C.
\newblock Transitive reduction for social network analysis and visualization.
\newblock In \emph{The 2005 IEEE/WIC/ACM International Conference on Web
  Intelligence (WI 2005)}. IEEE, pp. 128--131.

\bibitem{gansner2000open}
Gansner ER and North SC.
\newblock An open graph visualization system and its applications to software
  engineering.
\newblock \emph{Software: practice and experience} 2000; 30(11): 1203--1233.

\bibitem{burkhard2005tube}
Burkhard RA and Meier M.
\newblock Tube map visualization: Evaluation of a novel knowledge visualization
  application for the transfer of knowledge in long-term projects.
\newblock \emph{Journal of Universal Computer Science} 2005; 11(4): 473--494.

\bibitem{garcia2015visualisation}
Garc{\'\i}a F, Moraga M{\'A}, Serrano M et~al.
\newblock Visualisation environment for global software development management.
\newblock \emph{IET Software} 2015; 9(2): 51--64.

\bibitem{shamim2015evaluation}
Shamim A, Balakrishnan V and Tahir M.
\newblock Evaluation of opinion visualization techniques.
\newblock \emph{Information visualization} 2015; 14(4): 339--358.

\bibitem{wenskovitch2020interactive}
Wenskovitch J and North C.
\newblock Interactive artificial intelligence: Designing for the "two black
  boxes" problem.
\newblock \emph{Computer} 2020; 53(8): 29--39.
\newblock \doi{10.1109/MC.2020.2996416}.

\bibitem{seyser2018scrollytelling}
Seyser D and Zeiller M.
\newblock Scrollytelling – an analysis of visual storytelling in online
  journalism.
\newblock In \emph{2018 22nd International Conference Information Visualisation
  (IV)}. pp. 401--406.
\newblock \doi{10.1109/iV.2018.00075}.

\bibitem{hendricks2015constructing}
Hendricks RK and Boroditsky L.
\newblock Constructing mental time without visual experience.
\newblock \emph{Trends in Cognitive Sciences} 2015; 19(8): 429--430.
\newblock \doi{https://doi.org/10.1016/j.tics.2015.06.011}.

\bibitem{pherson2020structured}
Pherson RH and Heuer~Jr RJ.
\newblock \emph{Structured analytic techniques for intelligence analysis}.
\newblock CQ Press, 2020.

\bibitem{cheng2019explaining}
Cheng HF, Wang R, Zhang Z et~al.
\newblock Explaining decision-making algorithms through ui: Strategies to help
  non-expert stakeholders.
\newblock In \emph{Proc. of the 2019 CHI Conference on Human Factors in
  Computing Systems}. CHI ’19, New York, NY, USA: ACM.
\newblock ISBN 9781450359702, p. 1–12.

\bibitem{kang2010can}
Kang Ya, Görg C and Stasko J.
\newblock How can visual analytics assist investigative analysis? design
  implications from an evaluation.
\newblock \emph{IEEE Transactions on Visualization and Computer Graphics} 2011;
  17(5): 570--583.
\newblock \doi{10.1109/TVCG.2010.84}.

\bibitem{hughes2003discovery}
Hughes F and Schum D.
\newblock Discovery-proof-choice, the art and science of the process of
  intelligence analysis-preparing for the future of intelligence analysis.
\newblock \emph{Washington, DC: Joint Military Intelligence College} 2003; .

\bibitem{shneiderman2003eyes}
Shneiderman B.
\newblock The eyes have it: A task by data type taxonomy for information
  visualizations.
\newblock In Bederson BB and Shneiderman B (eds.) \emph{The Craft of
  Information Visualization}. Interactive Technologies, San Francisco: Morgan
  Kaufmann.
\newblock ISBN 978-1-55860-915-0, 2003.
\newblock pp. 364--371.
\newblock \doi{https://doi.org/10.1016/B978-155860915-0/50046-9}.

\end{thebibliography}



\end{document}